\documentclass[12pt]{article}
\usepackage{amsfonts}
\usepackage{amsmath}
\usepackage{amssymb}
\renewcommand{\theequation}{\arabic{equation}}

\def\PTP#1#2#3{Prog. Theor. Phys. {\bf #1} (#3), #2}

\def\PLB#1#2#3{Phys. Lett. B {\bf #1} (#3), #2}
\def\NPB#1#2#3{Nucl. Phys. B {\bf #1} (#3), #2}
\def\CMP#1#2#3{Commun. Math. Phys. {\bf #1} (#3), #2}
\def\MPLA#1#2#3{Mod. Phys. Lett. A {\bf #1} (#3), #2}
\def\PR#1#2#3{Phys. Rep. {\bf #1} (#3), #2}
\begin{document} 

\begin{flushright}
arXiv:0710.2959
\end{flushright}

\vspace{3mm}

\begin{center}
{\large \bf Galois Group of Elliptic Curves \\
                             and Flavor Symmetry }

\vspace{15mm}

Chuichiro HATTORI, 
            \footnote{E-mail: hattori@aitech.ac.jp} 
Mamoru MATSUNAGA,$^a$ 
            \footnote{E-mail: matsuna@phen.mie-u.ac.jp} \\
Takeo MATSUOKA,$^b$
            \footnote{E-mail: matsuoka@kogakkan-u.ac.jp}
and Kenichi NAKANISHI,$^c$ 
            \footnote{E-mail: nakanisi@bio.mie-u.ac.jp}
\end{center}

\begin{center}
\textit{
Science Division, General Education, Aichi Institute of Technology, \\
     Toyota 470-0392, JAPAN \\
{}$^a$Department of Physics Engineering, Mie University, \\
     Tsu 514-8507, JAPAN \\
{}$^b$Kogakkan University, Nabari 518-0498, JAPAN \\
{}$^c$Department of Sustainable Resource Science, Mie University, \\
     Tsu 514-8507, JAPAN }
\end{center}

\vspace{3mm}

\begin{abstract}
Putting emphasis on the relation between rational conformal 
field theory (RCFT) and algebraic number theory, 
we consider a brane configuration in which the D-brane 
intersection is an elliptic curve corresponding to RCFT. 
A new approach to the generation structure of fermions 
is proposed in which the flavor symmetry including the R-parity 
has its origin in the Galois group of elliptic curves with 
complex multiplication (CM). 
We study the possible types of the Galois group derived 
from the torsion points of the elliptic curve with CM . 
A phenomenologically viable example of the Galois group 
is presented, in which the characteristic texture of 
fermion masses and mixings is reproduced and 
the mixed-anomaly conditions are satisfied. 
\end{abstract}

\newpage 
\section{Introduction}

The generation structure of quarks and leptons forms 
the major outstanding problem of particle physics. 
In order to solve this problem, it is indispensable for us 
to introduce basic ingredients other than gauge symmetry.
The characteristic texture of fermion masses and mixings 
strongly suggests the existence of some kinds of underlying 
discrete symmetry. 
Furthermore, in the minimal supersymmetric standard model (MSSM) 
the additional discrete symmetry, that is, the R-parity ought 
to be introduced to forbid unfavorable interactions 
which bring about the fast proton decay. 
It is natural that the R-parity is combined with the usual flavor 
symmetry and that the flavor symmetry is interpreted 
in a broader sense including the R-parity. 
Up to now, many attempts have been made on the usual flavor 
symmetry. 
In these attempts, however, various types of the flavor symmetry 
are assumed to be temporarily applicable and individual analyses 
have been made as to whether the observed texture is derived from 
the assumed flavor symmetry. 
At present, no one has a first principle guide to single 
one symmetry over another. 
In this paper we direct our attention to the relation 
between string theory and algebraic number theory and 
explore a theoretically promising clue to the flavor symmetry 
in terms of arithmetic structure of the string theory.

In the string theory the flavor symmetry is considered to 
stem from the geometry of Calabi-Yau manifold on which 
strings propagate. 
As shown in the Gepner model,\cite{Gepner1} 
the geometry of Calabi-Yau manifold is closely linked to 
the 2-dimensional rational conformal field theory (RCFT). 
In addition, RCFT is also related to 
algebraic number field of finite degree. 
In fact, the fusion rule in RCFT\cite{Verlinde} 
yields some commutative ring which is isomorphic to ring of 
integers of algebraic number field determined by 
the fusion rule.\cite{Gepner2} 
Therefore, the geometry of Calabi-Yau manifold can be studied 
in terms of arithmetic structure of the algebraic number field. 
Furthermore, RCFT also connects with elliptic curves through 
their modular properties. 
Explicitly, when the $c = 2$ RCFT on two-torus $T^2$ is given, 
we find the corresponding elliptic curve. 
Although most elliptic curves over $\mathbb{C}$ have only 
the multiplication-by-$n$ $(n \in \mathbb{Z}_+)$ endomorphisms, 
the elliptic curve corresponding to the $c = 2$ RCFT possesses 
extra endomorphism called complex multiplication (CM).\cite{Gukov} 
Therefore, when a Calabi-Yau manifold contains two-torus $T^2$ 
given by the elliptic curve with CM as its submanifold, 
it is most likely that the arithmetic structure of 
the elliptic curve affects the 4-dimensional effective theory. 
The purpose of this paper is to study the flavor symmetry 
from string theory and its characteristics 
in terms of the arithmetic structure of 
the elliptic curve with CM.

Let us consider the lattice $\Lambda$ 
with periods $\omega_1$ and $\omega_2$ as 
\begin{equation}
  \Lambda = \{ \omega = \mathbb{Z} \,\omega _1  
                       + \mathbb{Z} \,\omega _2 \ | \ 
    \omega_1, \ \omega_2 \in \mathbb{C}, \ 
                       {\rm Im} (\omega_1/\omega_2) > 0 \}.
\end{equation}
The lattice $\Lambda$ defines an elliptic curve $E_{\Lambda}$ 
through the one-to-one complex analytic map 
\begin{equation}
\begin{array}{rcl}
  \mathbb{C}/{\Lambda} & \longrightarrow & E_{\Lambda} \ : \ 
         y^2 = 4 \, x^3 - g_2(\Lambda) \, x - g_3(\Lambda), \\
    u & \longmapsto & (x, \ y) = 
           (\wp(u; \ \Lambda), \ \wp'(u; \ \Lambda)), 
\end{array}
\label{eqn:map}
\end{equation}
where $\wp(u; \ \Lambda)$ is the Weierstrass $\wp$-function 
relative to the lattice $\Lambda$. 
The set of points of $\mathbb{C}/\Lambda$ forms an additive group. 
The map $\mathbb{C}/{\Lambda} \rightarrow E_{\Lambda}$ 
preserves its additive group structure and 
gives the isomorphism 
${\rm End}(\mathbb{C}/\Lambda ) \cong {\rm End}(E_{\Lambda})$. 
Elliptic curves $E_{\Lambda}$ are classified via the 
$\mathbb{C}$-isomorphism related to the homothety of 
the lattice $\Lambda$ and a $\mathbb{C}$-isomorphism class is 
denoted by $\{ E_{\Lambda} \}$. 
As a representative of $\{ E_{\Lambda} \}$ we can choose 
the elliptic curve determined by the normalized lattice 
$\Lambda _{\tau} = \{ \mathbb{Z}  + \mathbb{Z} \tau \}$, 
where $\tau = \omega_1/\omega_2$ is the complex structure 
parameter of the elliptic curve. 
If the elliptic curve $E_{\Lambda}$ has CM, then 
the complex structure parameter $\tau$ is a root of 
some quadratic equation with integer coefficients 
and the algebraic extension of $\mathbb{Q}$ associated with 
$E_{\Lambda}$, $K = \mathbb{Q}(\tau)$, becomes a quadratic 
imaginary field. 
In particular, in the $c = 2$ RCFT on $T^2$ the K\"ahlar 
modulus of the corresponding elliptic curve 
$E_{\Lambda}$ takes its value in $K$.\cite{Gukov} 
The elliptic curves over $\mathbb{C}$ are apparently smooth 
manifolds but those connected with RCFT conceal a peculiar 
discrete symmetry linked to the arithmetic structure. 
In order to explore such a discrete symmetry in RCFT, 
we addrress ourselves to the study of the 
arithmetic structure of the elliptic curve with CM.

The $j$-invariant $j(E_{\Lambda})$, which is the standard 
modulus of the elliptic curve $E_{\Lambda}$, 
is a function depending only on $\tau$. 
When the elliptic curve $E_{\Lambda}$ possesses CM, 
$j(E_{\Lambda})$ becomes an algebraic integer and 
degree of extension of the field 
$\mathbb{Q}(j(E_{\Lambda}))/\mathbb{Q}$ is equal to 
the class number of the quadratic imaginary field 
$K = \mathbb{Q}(\tau)$.\cite{Silverman} 
Furthermore, $\mathbb{Q}(j(E_{\Lambda}))$ is the minimal 
field of definition for the $\mathbb{C}$-isomorphism class 
$\{ E_{\Lambda} \}$. 
In this paper, we make the assumption 
\begin{equation}
  j(E_{\Lambda}) \in \mathbb{Q} 
\label{quantize}
\end{equation}
for the standard modulus. 
Then, $K$ has the class number 1 and 
the elliptic curve $E_{\Lambda}$ can be defined over $\mathbb{Q}$. 
This assumption implies a 
{\itshape quantization} 
for the shape of $E_{\Lambda}$. 
Specifically, in the diagonal RCFT the size of the extra space 
is also quantized. 
For instance, the diagonal RCFT on $S^1$ leads to 
the quantization condition $R^2 \in \mathbb{Z}$ 
or $R^{-2} \in \mathbb{Z}$, 
where $R$ is the radius of the circle. 
Similarly, in the diagonal RCFT on $E_{\Lambda}$ 
K\"ahlar modulus of $E_{\Lambda}$ becomes CM.\cite{Gukov}

In the context of nonperturbative string theory RCFT naturally 
leads to the D-brane (boundary state, Cardy state).\cite{Cardy} 
The D-brane is described in terms of a submanifold in 
the Calabi-Yau manifold corresponding to the RCFT.\cite{Gukov} 
In this paper we postulate that there exist two kinds of 
D-brane and that the intersection of these D-branes gives 
$M_4 \times T^2$, 
where $M_4$ represents 4-dimensional spacetime. 
The two-torus $T^2$ in the extra dimensional space 
is given by the elliptic curve $E_{\Lambda}$. 
In the D-brane picture the quark/lepton superfields and 
the Higgs superfields are equally described in terms of 
open strings with both ends attached to $E_{\Lambda}$. 
Although the endpoints of an open string are restricted to lie 
on the $E_{\Lambda}$(D-brane), they are free in the direction 
along the $E_{\Lambda}$.

We now assume that the extra space described in terms of RCFT 
is discretized and is a finite set of points on 
the elliptic curve $E_{\Lambda}$. 
This assumption is in line with the fact that 
RCFT is defined over the fusion variety which 
consists of points determined by the ideal of the polynomial 
ring isomorphic to the fusion ring.\cite{Gepner2} 
Concretely, the discrete extra space is taken as all of 
the $m$-torsion points 
\begin{equation}
   E_{\Lambda}[m] = \{ \ P \in E_{\Lambda} \ | \ m \,P = {\cal O} \} 
\label{m-torsion}
\end{equation}
of the elliptic curves $E_{\Lambda}$ with CM, 
where ${\cal O}$ is the zero element of additive group on 
$E_{\Lambda}$. 
In this paper we consider the case $m = p^e$ with a prime number 
$p$ and $e \in \mathbb{Z}_+$. 
$E_{\Lambda}[m]$ on which the both endpoints of an open string 
lie has the structure 
\begin{equation}
  E_{\Lambda}[m] \cong \frac{\mathbb{Z}}{m \mathbb{Z}} 
                       \oplus \frac{\mathbb{Z}}{m \mathbb{Z}}. 
\end{equation}
A point on $E_{\Lambda}[m]$ is expressed as $P_l = (x_l, \ y_l) = 
(\wp(u_l; \ \Lambda), \ \wp'(u_l; \ \Lambda))$ with 
$l = (l_1, \ l_2)$ and $u_l = ( l_1 \omega_1 + l_2 \omega_2 )/m$ 
where $l_1, \ l_2 = 0, \cdots m-1$.

Let us consider an orientable open string which corresponds 
to a matter superfield $\Phi_\nu$. 
Here the matter superfield $\Phi_\nu$ represents the quark/lepton 
superfield and the Higgs superfield classified by the generation. 
Since the string is free in the direction along the 
$E_{\Lambda}[m]$, 
the endpoint $P_l$ runs over all points of $E_{\Lambda}[m]$. 
Each point $P_l$ of $E_{\Lambda}[m]$ is accompanied by 
another endpoint $P_{l'}$ of $E_{\Lambda}[m]$. 
The matter superfield $\Phi_\nu$ is characterized by a set of 
the ordered pairs of their endpoints $(P_l, \ P_{l'})$ denoted by 
$\{ (P_l, \ P_{l'}) \}_{\nu}$. 
It is assumed that if both $(P_l, \ P_{l'})$ and $(P_k, \ P_{k'})$ 
are contained in the set $\{ (P_l, \ P_{l'}) \}_{\nu}$, 
the $p$-adic distances~\cite{Koblitz} of the both endpoints 
satisfy the condition 
\begin{equation}
  |P_l - P_k|_p =  |P_{l'} - P_{k'}|_p, 
\label{pdistance}
\end{equation}
where $|P_l|_p \equiv {\rm Max} \{ |l_1/m|_p, \ |l_2/m|_p \}$ 
and $|P_0|_p = |{\cal O}|_p = 0$. 
This condition bears some analogy with the parallel transport 
in the sense of the ordinary distance. 
From the additive group structure on $E_{\Lambda}[m]$, 
we have $P_l - P_k = P_{l-k}$. 
Then the above condition is rewritten as 
$|P_{l-k}|_p =  |P_{l'-k'}|_p$. 
If we take $l' = k' \ \pmod{m}$, we have 
$|P_{l-k}|_p = |P_0|_p = |{\cal O}|_p = 0$. 
This leads to $l = k \ \pmod{m}$. 
Consequently, in the set $\{ (P_l, \ P_{l'}) \}_{\nu}$ 
we may consider the mapping 
\begin{equation}
   \phi_{\nu} \ : \  P_l \ \mapsto \ P_{l'} \qquad 
      {\rm for \ all} \ P_l \in E_{\Lambda}[m]. 
\end{equation}
The mapping $\phi_{\nu}$ is a one-to-one map from 
$E_{\Lambda}[m]$ to $E_{\Lambda}[m]$.

Symmetry properties of $E_{\Lambda}[m]$ closely connect 
with those of the matter superfields. 
Elliptic curves $E_{\Lambda}$ have the multiplication-by-$n$ 
$(n \in \mathbb{Z}_+)$ endomorphism, 
which induces the endomorphism of $E_{\Lambda}[m]$. 
In particular, when $(n, \ m) = 1$, this endomorphism of 
$E_{\Lambda}[m]$ is an automorphism. 
It is natural that the matter superfield $\Phi_\nu$ is invariant 
under the multiplication-by-$n$ automorphism. 
Thus we expect the relation 
\begin{equation}
  \{ (P_{nl}, \ P_{nl'}) \}_{\nu}  \subset 
      \{ (P_l, \ P_{l'}) \}_{\nu} \qquad 
                   {\rm for} \ n \in \mathbb{Z}_+. 
\end{equation}
In the case $n=m$, $(P_{nl}, \ P_{nl'})$ amounts to 
$({\cal O}, \ {\cal O})$. 
If we take $P_k = P_{k'} = {\cal O}$ in Eq.~\eqref{pdistance}, 
we obtain $|P_l|_p =  |P_{l'}|_p$. 
This means that the order of $P_{l'}$ in the additive group 
coincides with that of $P_l$. 
It turns out that the mapping $\phi_{\nu}$ keeps 
the order of each point of $E_{\Lambda}[m]$.

On the other hand, the transformation of $P_l = (x_l, \ y_l)$ 
into $P_{l'} = (x_{l'}, \ y_{l'})$ 
induces the action on the Galois extension field $L$ 
which is the extension of field $K(j(E_{\Lambda}))$ generated by 
the coordinates of all of $E_{\Lambda}[m]$. 
We notice that $K(j(E_{\Lambda})) = K$ because of 
the above assumption~\eqref{quantize}.
It is expected that the string theory is described 
in terms of RCFT and that the RCFT is closely linked to 
the Galois extension field $L$. 
Thus, we hypothesize that the one-to-one and 
order-preserving mapping $\phi_{\nu}$ 
corresponds to a conjugation mapping of $L$. 
This means that the mapping $\phi_{\nu}$ is an element of 
the automorphism group of $L$. 
Hence we are led to the study of the Galois group of $L$.

The quadratic imaginary field $K/\mathbb{Q}$ leads to 
the Galois group $Gal(K/\mathbb{Q}) = \mathbb{Z}_2$. 
The field $L = K(E_{\Lambda}[m])$ is a Galois 
extension of $K$ and the Galois group 
\begin{equation}
   H = Gal (L/K) = {\rm Aut}_KL 
\end{equation}
is abelian.\cite{Silverman} 
In addition, $L$ is a Galois extension of $\mathbb{Q}$ 
and the Galois group 
\begin{equation}
   G = Gal (L/\mathbb{Q}) = {\rm Aut}_{\mathbb{Q}}L 
\end{equation}
is derived. 
Each element of the Galois group $G$ acts on $E_{\Lambda}[m]$. 
We thus obtain the Galois representation 
\begin{equation}
  \rho_m : Gal(L/\mathbb{Q}) 
               \rightarrow GL_2(\mathbb{Z}/m \mathbb{Z}), 
\end{equation}
which is a one-to-one homomorphism. 
Consequently, the degree of freedom of matter superfield 
connects with an element of the Galois group $G$. 
In our point of view the Galois group $G$ is precisely 
the flavor symmetry in the effective theory from string. 
In general, $L$ is not an abelian extension of $\mathbb{Q}$. 
Since $G$ is an extension of $\mathbb{Z}_2$ by the kernel $H$, 
in many cases $G$ contains the dihedral group $D_n \ (n \geq 3)$ 
but not the symmetric group $S_n \ (n \geq 4)$.

This paper is organized as follows. 
A brief explanation of the elliptic curves with CM 
is given in section 2. 
In section 3 we proceed to study the main features of 
the Galois group $G$ derived from the $m$-torsion 
points $E_{\Lambda}[m]$ $(m \in \mathbb{Z}_+)$ of 
the elliptic curves $E_{\Lambda}$ with CM. 
We postulate that there exist two kinds of 
D-brane and that the intersection of these D-branes gives 
$M_4 \times T^2$, where the $T^2$ is given by $E_{\Lambda}$. 
The quantization condition on the shape of $E_{\Lambda}$ 
is introduced through the assumption 
$j(E_{\Lambda}) \in \mathbb{Q}$. 
We propose a new interpretation that the flavor symmetry including 
the R-parity has its origin in the Galois group $G$ which 
stems from the arithmetic structure of elliptic curves 
with CM linked to RCFT. 
Concrete examples of the Galois group $G$ are also shown 
for the cases $\tau = i, \ \omega$, where $\omega = \exp (2\pi i/3)$. 
In section 4 we discuss the group extension $G$ of $\mathbb{Z}_2$ 
by an abelian kernel $H$ and classify possible Galois groups $G$ 
in which $\mathbb{Z}_2$  is homomorphically embedded in $G$. 
In section 5 we apply the Galois group 
$G = \{ \mathbb{Z}_2 \ltimes 
  \, (\mathbb{Z}_4 \times \mathbb{Z}'_4) \} \times \mathbb{Z}_N$ 
to $SU(6) \times SU(2)_R$ string-inspired model 
as an example of the flavor symmetry. 
$\mathbb{Z}_2 = Gal(K/\mathbb{Q})$ is identified with the R-parity. 
We take the D-brane configuration in which 
one of two kinds of the D-brane has 
the degree of freedom of $SU(6)$ gauge group and 
the other has that of $SU(2)_R$ gauge group. 
This example exhibits a phenomenologically viable solution 
with $N = 31$ in which the characteristic texture of 
fermion masses and mixings is reproduced. 
Furthermore, it is also shown that this solution satisfies 
the mixed-anomaly conditions. 
Section 6 is devoted to summary and discussion. 
In Appendix we consider the general classification of 
the group extension of $\mathbb{Z}_2$ by an abelian group $H$.

\vspace{5mm}


\section{Elliptic curves with CM}

In this section we give a brief explanation of 
the elliptic curve $E$ and CM on $E$. 
After that, we will discuss the effective theory in which 
the quark/lepton superfields and the Higgs superfields 
live in $M_4 \times T^2$, where the $T^2$ is given 
by $E$ corresponding to RCFT. 
When an elliptic curve $E$ corresponding to a certain RCFT 
is given, as a consequence of its rationality 
the elliptic curve $E$ possesses CM.\cite{Gukov}

The elliptic curve $E_{\Lambda}$ relative to the lattice 
$\Lambda$ with periods $\omega_1$ and $\omega_2$ is defined 
via the one-to-one complex analytic map 
\begin{equation}
\begin{array}{rcl}
  \mathbb{C}/{\Lambda} & \longrightarrow & E_{\Lambda} \ : \ 
         y^2 = 4 \, x^3 - g_2(\Lambda) \, x - g_3(\Lambda), \\
    u & \longmapsto & (x, \ y) = 
           (\wp(u; \ \Lambda), \ \wp'(u; \ \Lambda)). 
\end{array}
\label{eqn:map}
\end{equation}
The functions $g_2$ and $g_3$ are expressed in terms of 
Eisenstein series $G_{2k}(\Lambda)$ as 
\begin{equation}
  g_2(\Lambda) = 60 \, G_{4}(\Lambda), \qquad 
  g_3(\Lambda) = 140 \, G_{6}(\Lambda), \nonumber 
\end{equation}
where 
\begin{equation}
  G_{2k}(\Lambda) = \sum_{\substack{\omega \in \Lambda \\
      \omega \neq 0}} \frac{1}{\omega^{2k}}. \nonumber 
\end{equation}
Substituting $2y$ for $y$ and using the notation 
$g_2(\Lambda) = -4A$ and $g_3(\Lambda) = -4B$, 
we obtain the standard form of the elliptic curve 
\begin{equation}
     E_{\Lambda} \ : \ \ y^2 = f(x) = x^3 + A x + B. 
\label{Elambda}
\end{equation}
The modular discriminant of the elliptic curve $E_{\Lambda}$ is 
given by 
\begin{equation}
   \Delta (\Lambda) = g_2(\Lambda)^3 - 27 \, g_3(\Lambda)^2 
                 = -16 \, \left( 4 \, A^3 + 27 \, B^2 \right). 
                 \nonumber 
\end{equation}
The set of points of an elliptic curve forms an additive group 
and the group law is given by rational functions. 
The map $\mathbb{C}/{\Lambda} \rightarrow E_{\Lambda}$ defined 
by Eq.~(\ref{eqn:map}) preserves its additive group structure and 
gives the isomorphism 
${\rm End}(\mathbb{C}/\Lambda ) \cong {\rm End}(E_{\Lambda})$. 
When we carry out the homothety of $\Lambda$ 
\begin{equation}
  \Lambda \longrightarrow \lambda \Lambda, 
                         \qquad \lambda \in \mathbb{C}^*, 
\end{equation}
the coefficients $A$ and $B$ are transformed as 
\begin{equation}
   A \longrightarrow \lambda^{-4} A, \qquad 
                  B \longrightarrow \lambda^{-6} B. \nonumber 
\end{equation}
If we substitute $(\lambda^{-2} x, \ \lambda^{-3} y)$ for 
$(x, \ y)$, Eq.~\eqref{Elambda} remains unchanged. 
For this reason, elliptic curves $E_{\Lambda}$ are 
classified via the $\mathbb{C}$-isomorphism. 
As a representative of a $\mathbb{C}$-isomorphism class 
$\{ E_{\Lambda} \}$ we can choose $E_{\Lambda_{\tau}}$ 
relative to the normalized lattice 
$\Lambda _{\tau} = \{ \mathbb{Z}  + \mathbb{Z} \tau \}$ 
with $\tau = \omega_1/\omega_2$. 
Here, the complex structure parameter $\tau$ of two-torus $T^2$ 
is defined in the upper half-plane 
$U = \{ \tau \in \mathbb{C} \ | \ {\rm Im} \, \tau > 0 \}$. 
However, many $\tau$'s give the same lattice $\Lambda_{\tau}$. 
This is because there remains the degree of freedom of modular 
transformations. 
Hence, we introduce the quotient space 
$\Gamma(1) \backslash U^*$ given by the fundamental region 
\begin{equation}
  {\cal F} = \{ \tau \in U \ | \ |\tau| \geq 1, \ 
                     |{\rm Re} \, \tau | \leq \frac{1}{2} \}, 
\end{equation}
where $\Gamma (1)$ is the modular group and 
$U^* = U \cup \mathbb{ P}^1(\mathbb{Q})$. 
If we take the value of $\tau$ in ${\cal F}$, 
$\tau$ has the one-to-one correspondences 
to the lattice $\Lambda_{\tau}$. 
The functions $g_2(\tau) := g_2(\Lambda_{\tau})$ and 
$g_3(\tau) := g_3(\Lambda_{\tau})$ are the modular 
forms of weight 4 and 6 for $\Gamma (1)$, respectively. 
The modular discriminant $\Delta (\tau)$ is the cusp form 
of weight 12 and expressed as 
\begin{equation}
   \Delta (\tau) = g_2(\tau)^3 - 27 \, g_3(\tau)^2 
                 = (2 \pi)^{12} \eta(\tau)^{24}, \nonumber 
\end{equation}
where $\eta(\tau)$ stands for the Dedekind $\eta$-function.

Next we proceed to explain CM on the elliptic curve. 
In general, using the isomorphism 
${\rm End}(E_{\Lambda}) \cong {\rm End}(\mathbb{C}/\Lambda )$, 
we can study the automorphism on $\mathbb{C}/\Lambda$ 
instead of that on $E_{\Lambda}$. 
The elliptic curve corresponding to RCFT possesses 
not only the multiplication-by-$n$ ($n \in \mathbb{Z}_+$) 
endomorphism but also CM. 
Namely, there exists a complex number $\mu \ (\not\in \mathbb{ R})$ 
such that $\mu \Lambda \subset \Lambda$. 
In the elliptic curve $E_{\Lambda}$ with CM, 
$\tau$ is a root of some quadratic equation with integer 
coefficients. 
The discriminant of the quadratic equation $D$ should be 
negative. 
Thus we can define the quadratic imaginary field 
$K = \mathbb{Q}(\sqrt{D}) = \mathbb{Q}(\tau)$, which is 
the algebraic extension of $\mathbb{Q}$ associated with 
$E_{\Lambda}$. 
This means that $E_{\Lambda}$ possesses CM by $K$. 
In the $c = 2$ RCFT on $T^2$ the K\"ahlar modulus 
of $E_{\Lambda}$ also takes its value in $K$.

The $j$-invariant $j(E_{\Lambda})$ depends only upon $\tau$ 
and so is denoted by $j(\tau)$. 
The explicit form is 
\begin{equation}
   j(\tau) = \frac{ \left( 12 \, g_2(\tau) \right)^3 }{ \Delta (\tau) }
           = 1728 \ \frac{4 A^3}{4 A^3 + 27 B^2}, \nonumber 
\end{equation}
which is a modular function of weight 0. 
For the elliptic curve with CM, $j(E_{\Lambda})$ becomes 
an algebraic integer and $\mathbb{Q}(j(E_{\Lambda}))$ is 
the minimal field of definition for the $\mathbb{C}$-isomorphism 
class $\{ E_{\Lambda} \}$. 
The class $\{ E_{\Lambda} \}$ is isomorphic to the ideal 
class group of $R_K$ which is the ring of integers of $K$. 
The degree of extension of the field 
$\mathbb{Q}(j(E_{\Lambda}))/\mathbb{Q}$ is equal to 
the class number $h_K$ of the quadratic imaginary field 
$K = \mathbb{Q}(\tau)$.\cite{Silverman}

Here, we make the assumption 
\begin{equation}
  j(E_{\Lambda}) \in \mathbb{Q} \,. 
\end{equation}
This means $h_K = 1$. 
Hereafter, we refer this assumption to as 
{\itshape quantum hypothesis} for the $j$ function. 
This hypothesis strongly constrains the value of $\tau$ 
and then implies the quantization for the shape of $E_{\Lambda}$. 
In this case of $E_{\Lambda}$ the coefficients in the standard 
form $f(x)$ in Eq.~\eqref{Elambda} can be taken as 
\begin{equation}
   A, \ B \in \mathbb{Z}. \nonumber 
\end{equation}
Incidentally, it is known that there are only nine 
quadratic imaginary fields $\mathbb{Q}(\sqrt{D})$ of 
$h_K = 1$.\cite{Silverman} 
These fields are the cases of 
\begin{equation}
-D = 1, \ 2, \ 3, \ 7, \ 11, \ 19, \ 43, \ 67, \ 163. 
\end{equation}
In the next section, we take up the cases $-D = 1, \ 3$ 
as simple examples of the Galois group of the elliptic 
curves with CM.

\vspace{5mm}


\section{Galois groups on elliptic curves}

In our approach based on the D-brane picture, 
the quark/lepton superfields and the Higgs superfields are 
described in terms of open strings with both ends 
attached to the extra space $E_{\Lambda}$. 
These endpoints of open strings are free in the direction 
along the $E_{\Lambda}$. 
In this paper the extra space is assumed to be all of 
the $m$-torsion points $E_{\Lambda}[m]$ of the elliptic 
curves $E_{\Lambda}$ with CM. 
Thus the endpoints of open strings lie on $E_{\Lambda}[m]$. 
A point on $E_{\Lambda}[m]$ is expressed as $P_l = (x_l, \ y_l)$ 
with $l = (l_1, \ l_2)$. 
Since the string is free in the direction along the 
$E_{\Lambda}[m]$, 
the endpoint $P_l$ runs over all points of $E_{\Lambda}[m]$. 
Each point $P_l$ of $E_{\Lambda}[m]$ is accompanied by 
another endpoint $P_{l'}$ of $E_{\Lambda}[m]$. 
The matter superfield $\Phi_\nu$ is characterized by a set of 
the ordered pairs of their endpoints $(P_l, \ P_{l'})$ denoted by 
$\{ (P_l, \ P_{l'}) \}_{\nu}$. 
In conjunction with this set $\{ (P_l, \ P_{l'}) \}_{\nu}$ we 
consider the mapping 
\begin{equation}
   \phi_{\nu} \ : \  P_l \ \mapsto \ P_{l'} \qquad 
      {\rm for \ all} \ P_l \in E_{\Lambda}[m]. 
\end{equation}
It is assumed that if both $(P_l, \ P_{l'})$ and $(P_k, \ P_{k'})$ 
are contained in the set $\{ (P_l, \ P_{l'}) \}_{\nu}$, 
the $p$-adic distances of the both endpoints satisfy 
the condition $|P_l - P_k|_p =  |P_{l'} - P_{k'}|_p$. 
As mentioned in \S 1, it follows that the mapping 
$\phi_{\nu}$ is a one-to-one map from $E_{\Lambda}[m]$ to 
$E_{\Lambda}[m]$.

Elliptic curves $E_{\Lambda}$ have the multiplication-by-$n$ 
$(n \in \mathbb{Z}_+)$ endomorphisms. 
In particular, when $(n, \ m) = 1$, $E_{\Lambda}[m]$ has 
the multiplication-by-$n$ automorphism. 
In view of the close relation between the symmetry property 
of $E_{\Lambda}[m]$ and that of the matter superfields, 
it is natural that the matter superfield $\Phi_\nu$ is invariant 
under the multiplication-by-$n$ automorphism. 
Thus we assume the relation 
$\{ (P_{nl}, \ P_{nl'}) \}_{\nu} \subset \{ (P_l, \ P_{l'}) \}_{\nu}$ 
for $n \in \mathbb{Z}_+$. 
Taking $(P_{ml}, \ P_{ml'}) = ({\cal O}, \ {\cal O})$ into account, 
we obtain $|P_l|_p =  |P_{l'}|_p$. 
It turns out that the mapping $\phi_{\nu}$ retains the order 
of the endpoint.

On the other hand, $E_{\Lambda}[m]$ equips the Galois 
extension field $L$. 
In fact, $L$ is the extension of field $K(j(E_{\Lambda}))$ 
generated by the coordinates of all of $E_{\Lambda}[m]$. 
The transformation of $P_l = (x_l, \ y_l)$ into 
$P_{l'} = (x_{l'}, \ y_{l'})$ acts on the Galois extension 
field $L$. 
It is expected that the string theory is described 
in terms of RCFT and that the RCFT is closely linked to 
the Galois extension field $L$. 
From this point of view and from the fact that $\phi_{\nu}$ 
is one-to-one and order-preserving, we hypothesize that 
$\phi_{\nu}$ corresponds to a conjugation mapping of $L$. 
This means that the mapping $\phi_{\nu}$ is an element of 
the automorphism group of $L$. 
Thus we are led to the study of the the Galois group of $L$. 
According to the hypothesis $j(E_{\Lambda}) \in \mathbb{Q}$ 
we have $K(j(E_{\Lambda})) = K$. 
The Galois extension field $L = K(E_{\Lambda}[m])$ has 
many important features. 
Concretely, the Galois group 
\begin{equation}
   H = Gal (L/K) = {\rm Aut}_KL 
\end{equation}
is abelian. 
Furthermore, $L$ is also the Galois extension of $\mathbb{Q}$ 
and its Galois group 
\begin{equation}
   G = Gal (L/\mathbb{Q}) = {\rm Aut}_{\mathbb{Q}}L 
\end{equation}
is the group extension of $\mathbb{Z}_2 = Gal (K/\mathbb{Q})$ 
by an abelian kernel $H$, as noted in section 4. 
In general, $L$ is not an abelian extension of $\mathbb{Q}$. 
Each element of the Galois group $G$ acts on $E_{\Lambda}[m]$. 
We obtain the Galois representation 
\begin{equation}
  \rho_m : Gal(L/\mathbb{Q}) 
               \rightarrow GL_2(\mathbb{Z}/m \mathbb{Z}). 
\end{equation}
Our hypothesis implies that the map $\phi_{\nu}$ represents 
an element of $G$. 
Consequently, the Galois group $G$ turns out to be precisely 
the flavor symmetry in the effective theory from string.

We are in a position to give a physical interpretation 
of $E_{\Lambda}[m]$. 
We may consider that the extra space $E_{\Lambda}[m]$ 
is quantized in unit of the size given by the fundamental 
period-parallelogram $\Lambda_0 = m^{-1}\Lambda$. 
Thus $m$ represents the size of the extra space 
$E_{\Lambda}[m]$ in unit of $E_{\Lambda_0}$. 
The K\"ahlar modulus $\rho$ of the $E_{\Lambda}$ becomes 
$\rho = m^2 \rho_0$, where $\rho_0$ is 
the K\"ahlar modulus of $E_{\Lambda_0}$. 
It is natural that the size of $E_{\Lambda_0}$ corresponds to 
the fundamental scale of the string theory, which is 
nearly equal to the Planck scale.

Among the $m$-torsion points on $\mathbb{C}/\Lambda$, 
the number of the points of order $m$ is denoted by $N_m$. 
The set of the points of order $m$ is given by 
\begin{equation}
 \{ u \ | \ u \in \mathbb{C}/\Lambda, \ m u \in \Lambda, \ 
               m' u \not\in \Lambda \ {\rm for} \ m' < m \}. 
               \nonumber 
\end{equation}
Any elements of $H$ give the mapping from the points of 
order $m$ to those of order $m$ on $\mathbb{C}/\Lambda$. 
In the case $m = p^e$ with a prime number $p$ 
we have 
\begin{equation}
 N_m =  ( p^2 - 1) \, p^{2 (e - 1)}. 
\end{equation}
In the case $m = 2$ we get $N_m = 3$, otherwise $N_m =$ even. 
The $x$-coordinates of the points of order $m$ 
among the $m$-torsion points on $E_{\Lambda}$ are 
the roots of the equation over $\mathbb{Q}$ 
\begin{equation}
 \psi_m(x) = \prod_u \, ( x - \wp(u)) = 0. 
\end{equation}
Since the Weierstrass $\wp$-function is an even function, 
the above product is taken over 3 points of order 2 on 
$\mathbb{C}/\Lambda$ for $m = 2$, otherwise over $N_m/2$ points. 
Therefore, the degree of the polynomial $\psi_m(x)$ 
is 3 for $m = 2$, otherwise $N_m/2$.

When $j$-invariant and $m$ are given, 
the Galois group $H$ of $L/K$ changes with changing 
the elliptic curve $E_{\Lambda}$. 
The maximum value of the order $\sharp H$ of $H$ is $N_m$. 
In order that the order $\sharp H$ of $H$ takes its maximum 
value, the polynomial $\psi_m(x)$ is needed to be irreducible 
over $\mathbb{Q}$. 
In general, if the polynomial $\psi_m(x)$ is irreducible over 
$\mathbb{Q}$, we have the relation $(\sharp H)|N_m$. 
On the other hand, if the polynomial $\psi_m(x)$ is 
reducible over $\mathbb{Q}$, 
$\sharp H$ is smaller than $N_m$ and is not always 
a divisor of $N_m$.

As concrete examples of the elliptic curves with CM 
we now consider two cases $-D = 1, \ 3$ which mean 
$\tau = i, \ \omega$, where $\omega = \exp (2\pi i/3)$. 
For these cases we illustrate the Galois group $G$ coming 
from $E_{\Lambda}[m]$. 
Since $g_3(i) = 0$ and $g_2(\omega) = 0$, 
the value of $j(E_{\Lambda})$ is 1728 and 0, respectively, 
that is, $j \in \mathbb{Q}$. 
Then $E_{\Lambda}$ associated with $\tau = i, \ \omega$ 
is $\mathbb{C}$-isomorphic to the elliptic curve 
$y^2 = x^3 + x$ and $y^2 = x^3 + 1$, respectively. 
Here we focus on the elliptic curves over $\mathbb{Q}$ 
\begin{eqnarray}
    E_{\Lambda} \ & : & \ y^2 = x^3 + A \, x, \ 
                              \qquad A \in \mathbb{Z}, \\
    E_{\Lambda} \ & : & \ y^2 = x^3 + B, \quad 
                              \qquad B \in \mathbb{Z},
\end{eqnarray}
which is $\mathbb{C}$-isomorphic to the elliptic curve 
$y^2 = x^3 + x$ and $y^2 = x^3 + 1$, respectively.

\begin{table}[t]
\caption{$m = 3$ torsion points on elliptic curve 
$E_{\Lambda} \ : \ y^2 = x^3 + 1$ }
\label{table:1}
\begin{center}
\begin{tabular}{|c|c|c|c|} \hline  \hline
\vphantom{\LARGE I}  &  $q = 0$   &   1   &   2  \\ \hline \hline
\vphantom{\LARGE I} $p = 0$  &  ${\cal O}$   
                     & $(-\sqrt[3]{4}, \ \sqrt{3} \,i)$ 
                     & $(-\sqrt[3]{4}, \ -\sqrt{3} \,i)$  \\ \hline
\vphantom{\LARGE I} 1   &  $(0, \ 1)$ 
          & \ \ $(-\sqrt[3]{4} \,\omega, \ \sqrt{3} \,i)$ \ \ 
          & \ \ $(-\sqrt[3]{4} \,\omega^2, \ -\sqrt{3} \,i)$ \ \ \\ \hline
\vphantom{\LARGE I} 2   & \phantom{M} $(0, \ -1)$ \phantom{M} 
            & $(-\sqrt[3]{4} \,\omega^2, \ \sqrt{3} \,i)$ 
            & $(-\sqrt[3]{4} \,\omega, \ -\sqrt{3} \,i)$   \\ \hline
\end{tabular}
\end{center}
\end{table}

For illustration we first take up the 3-torsion points on 
the elliptic curve $E_{\Lambda} : \ y^2 = x^3 + 1$ and 
study the Galois group $Gal(K(E_{\Lambda}[3])/\mathbb{Q})$. 
All of the 3-torsion points are described in terms of 
two bases $P$ and $Q$ as 
\begin{equation}
   E_{\Lambda}[3] = \{ p \, P + q \, Q \ | \ 
               p, \ q \in \mathbb{Z}/3 \mathbb{Z} \}. 
\end{equation}
Consequently, the torsion points are exhibited by 
2-dimensional representation of the Galois group. 
In the present case the coordinates $(x, \ y)$ of 
the bases $P$ and $Q$ are given by 
\begin{equation}
  P = (0, \ 1), \ \ \ \ Q = (- \sqrt[3]{4}, \ \sqrt{3}i). 
  \nonumber 
\end{equation}
All of the 3-torsion points are shown in Table 1. 
The Galois extension field becomes 
$L = K(E_{\Lambda}[3]) = 
        \mathbb{Q}(\omega, \ \sqrt[\scriptstyle 3]{4})$ 
and the Galois group of $L/K$ is $H = \mathbb{Z}_3$. 
Although $N_m = 8$ for the case $m = 3$, 
the polynomial $\psi_3(x)$ for $E_{\Lambda} : \ y^2 = x^3 + 1$ 
is reducible over $\mathbb{Q}$. 
This results in $\sharp H = 3$ and then $(\sharp H) \not| N_m$. 
The generators $b$ and $a$ of the Galois group $G$ are 
defined by 
\begin{eqnarray}
  b   \ & : & \ \omega \rightarrow \omega^2, \nonumber \\
  a \ & : & \ \sqrt[3]{4} \rightarrow \sqrt[3]{4}\omega. \nonumber 
\end{eqnarray}
Taking the Galois representation 
$\rho_3 : Gal(K(E_{\Lambda}[3])/\mathbb{Q}) 
               \rightarrow GL_2(\mathbb{Z}/3 \mathbb{Z})$, 
we have 
\begin{equation}
  \rho_3(b) \equiv \left(
   \begin{array}{cc}
       1  &  0  \\
       0  & -1  
   \end{array}
   \right), \ \ \ \ 
  \rho_3(a) \equiv \left(
   \begin{array}{cc}
       1  &  1  \\
       0  &  1  
   \end{array}
   \right) \ \ \ \pmod{3}. \nonumber 
\end{equation}
In this case we obtain the relations 
\begin{equation}
  b^2 = a^3 = e, \ \ \ \ b \, a \, b = a^{-1} 
\end{equation}
and the Galois group becomes the dihedral group 
\begin{equation}
 G = Gal(\mathbb{Q}(E_{\Lambda}[3])/\mathbb{Q}) 
               = \mathbb{Z}_2 \ltimes \mathbb{Z}_3 = D_3. 
\end{equation}

\begin{table}[t]
\caption{$m = 3$ torsion points on elliptic curve 
$E_{\Lambda} \ : \ y^2 = x^3 + x$ }
\label{table:2}
\begin{center}
\begin{tabular}{|c|c|c|c|} \hline  \hline
\vphantom{\LARGE I}   &  $q = 0$  &   1  &   2   \\ \hline \hline
\vphantom{\LARGE I} $p = 0$ &  ${\cal O}$   
                     & $(-\alpha_-, \ i \beta_{1-} )$ 
                     & $(-\alpha_-, \ -i \beta_{1-} )$   \\ \hline
\vphantom{\LARGE I} 1   & $(\alpha_-, \ \beta_{1-})$  
        & $(-i \alpha_+, \ \frac{(1+i)}{\sqrt{2}} \, \beta_{1+} )$ 
        & $( i \alpha_+, \ \frac{(1-i)}{\sqrt{2}} \, \beta_{1+} )$  \\ \hline
\vphantom{\LARGE I} 2   &  $(\alpha_-, \ -\beta_{1-})$  
        & $( i \alpha_+, \ \frac{(-1+i)}{\sqrt{2}} \, \beta_{1+} )$ 
        & $(-i \alpha_+, \ \frac{(-1-i)}{\sqrt{2}} \, \beta_{1+} )$ \\ \hline
\end{tabular}
\end{center}
\end{table}

As the second example we take up the 3-torsion points on 
the elliptic curve $E_{\Lambda} : \ y^2 = x^3 + x$ and 
study the Galois group $Gal(K(E_{\Lambda}[3])/\mathbb{Q})$. 
In the present case the coordinates $(x, \ y)$ of 
the bases $P$ and $Q$ are given by 
\begin{equation}
  P = (\alpha_-, \ \beta_{1-}), \qquad 
               Q = (-\alpha_-, \ i \beta_{1-}). \nonumber 
\end{equation}
In Table 2 all of the 3-torsion points are found, 
where 
\begin{equation}
   \alpha_{\pm}  = \frac{\sqrt{3} \pm 1}{\sqrt{2 \sqrt{3}}}, \qquad 
   \beta_{1\pm}  = \sqrt{\sqrt{3} \pm 1} \, 
          \left( 2/\sqrt{27} \right)^{\frac{1}{4}}. \nonumber 
\end{equation}
$\alpha_{\pm}$ and $\beta_{1+}/\sqrt{2}$ are described 
in terms of $\beta_{1-}$ as 
\begin{eqnarray}
  \alpha_+ & = & \frac{2}{3 \, \beta_{1-}^2},   \nonumber \\ 
  \alpha_- & = & \frac{2}{3 \, \beta_{1+}^2} = 
   \frac{3}{16} \, \beta_{1-}^2 (3 \, \beta_{1-}^4 + 4), \nonumber \\
  \frac{1}{\sqrt{2}} \, \beta_{1+} & = & 
         \frac{8 \, \beta_{1-}}{9 \, \beta_{1-}^4 + 4}. \nonumber 
\end{eqnarray}
Thus, the Galois extension field is 
$L = K(E_{\Lambda}[3]) = \mathbb{Q}(i, \ \beta_{1-})$ and 
the Galois group of $L/K$ becomes $H = \mathbb{Z}_8$. 
In this case the polynomial $\psi_3(x)$ is irreducible 
over $\mathbb{Q}$. 
The generators $b$ and $a$ of the Galois group $G$ are defined by 
\begin{eqnarray}
  b   \ & : & \ i \longrightarrow -i, \nonumber \\
  a \ & : & \ \beta_{1-} \longrightarrow 
              \frac{(-1+i)}{\sqrt{2}} \, \beta_{1+}. \nonumber 
\end{eqnarray}
Taking the Galois representation 
$\rho_3 : Gal(K(E_{\Lambda}[3])/\mathbb{Q}) 
             \rightarrow GL_2(\mathbb{Z}/3 \mathbb{Z})$, 
we have 
\begin{equation}
  \rho_3(b) \equiv \left(
   \begin{array}{cc}
       1  &  0  \\
       0  & -1  
   \end{array}
   \right), \ \ \ \ 
  \rho_3(a) \equiv \left(
   \begin{array}{cc}
       -1  &  -1  \\
        1  &  -1  
   \end{array}
   \right) \ \ \ \pmod{3}. \nonumber 
\end{equation}
These generators hold the relations 
\begin{equation}
  b^2 = a^8 = e, \qquad 
         b \, a \, b = a^3. 
\end{equation}
The Galois group $G$ is of the form 
\begin{equation}
 G = Gal(\mathbb{Q}(E_{\Lambda}[3])/\mathbb{Q}) 
               = \mathbb{Z}_2 \ltimes \mathbb{Z}_8. 
\end{equation}
Although this group differs from $D_8$, 
the relation 
\begin{equation}
   b \, a^2 \, b = a^{-2}
\end{equation}
is satisfied and then we obtain 
$\langle b, \ a^2 \rangle = 
D_4 \subset G = \mathbb{Z}_2 \ltimes \mathbb{Z}_8 $.

\begin{table}[t]
\caption{Galois Groups on $E_{\Lambda} \ : \ y^2 = x^3 + A \, x$}
\label{table:3}
\begin{center}
\begin{tabular}{|c|c|c|} \hline \hline
\vphantom{\LARGE I} \phantom{MMMM} $E_{\Lambda}$ \phantom{MMMM}  & 
\phantom{M} $y^2 = x^3 + x$      \phantom{M}  & 
\phantom{M} $y^2 = x^3 + 2x$     \phantom{M}    \\ \hline \hline
\vphantom{\LARGE I} $L = K(E_{\Lambda}[2])$ & 
    $\mathbb{Q}(i)$ & 
    $\mathbb{Q}(\sqrt{2}\,i)$  \\
\vphantom{\LARGE I} $G= Gal (L/\mathbb{Q})$  &
    $\mathbb{Z}_2$  &  $\mathbb{Z}_2$  \\ \hline
\vphantom{\LARGE I} $L = K(E_{\Lambda}[3])$ & 
    $\mathbb{Q}(i, \ \beta_{1-})$ & 
    $\mathbb{Q}(i, \ \beta_{2-})$  \\
\vphantom{\LARGE I} $G= Gal (L/\mathbb{Q})$  &
    $\mathbb{Z}_2 \ltimes \, \mathbb{Z}_8$  &  
    $\mathbb{Z}_2 \ltimes \, \mathbb{Z}_8$   \\ \hline
\vphantom{\LARGE I} $L = K(E_{\Lambda}[4])$ & 
    $\mathbb{Q}(i, \ \sqrt{2})$ & 
    $\mathbb{Q}(i, \ \sqrt[\scriptstyle 4]{2})$  \\
\vphantom{\LARGE I} $G= Gal (L/\mathbb{Q})$  &
    $\mathbb{Z}_2 \times \mathbb{Z}_2$  &  $D_4$  \\ \hline
\end{tabular}
\end{center}
\end{table}

\begin{table}[t]
\caption{Galois Groups on $E_{\Lambda} \ : \ y^2 = x^3 + B$}
\label{table:4}
\begin{center}
\begin{tabular}{|c|c|c|} \hline \hline
\vphantom{\LARGE I} \phantom{MMMM} $E_{\Lambda}$ \phantom{MMMM}  &
\phantom{Mi} $y^2 = x^3 + 1$      \phantom{Mi}  &
\phantom{Mi} $y^2 = x^3 + 2$      \phantom{Mi}   \\ \hline \hline
\vphantom{\LARGE I} $L = K(E_{\Lambda}[2])$ & 
    $\mathbb{Q}(\omega)$ & 
    $\mathbb{Q}(\omega, \ \sqrt[\scriptstyle 3]{2})$  \\
\vphantom{\LARGE I} $G= Gal (L/\mathbb{Q})$  &  
    $\mathbb{Z}_2$  &  $D_3$  \\ \hline
\vphantom{\LARGE I} $L = K(E_{\Lambda}[3])$ & 
    $\mathbb{Q}(\omega, \ \sqrt[\scriptstyle 3]{4})$  & 
    $\mathbb{Q}(\omega, \ \sqrt{2})$  \\
\vphantom{\LARGE I} $G= Gal (L/\mathbb{Q})$  &  $D_3$  &  
        $\mathbb{Z}_2 \times \mathbb{Z}_2$  \\ \hline
\vphantom{\LARGE I} $L = K(E_{\Lambda}[4])$ & 
    $\mathbb{Q}(\omega, \ \sqrt[\scriptstyle 4]{27}/\sqrt{2})$  & 
    $\mathbb{Q}(\omega, \ \sqrt[\scriptstyle 3]{2}, \ 
                         \sqrt[\scriptstyle 4]{27})$ \\
\vphantom{\LARGE I} $G= Gal (L/\mathbb{Q})$  &  $D_4$  & 
        $\mathbb{Z}_2 \ltimes \, 
      (\mathbb{Z}_2 \times \mathbb{Z}_2 \times \mathbb{Z}_3) $ \\ \hline
\end{tabular}
\end{center}
\end{table}

In Tables 3 and 4 we summarize the Galois extension fields 
$L = K( E_{\Lambda}[m] )$ and the Galois group $G = Gal (L/\mathbb{Q})$ 
for the cases $m = 2, \ 3, \ 4$, where the elliptic curves 
$E_{\Lambda}$ are taken as $y^2 = x^3 + Ax$ and $y^2 = x^3 + B$ 
with $A, \ B = 1, \ 2$, respectively. 
In Table 3 we use the notation $\beta_{2-} = \beta_{1-}/2^{1/4}$. 
In particular, it is worth noting that in the $4$-torsion 
points on $E_{\Lambda} \ : \ y^2 = x^3 + 2$ 
the polynomial $\psi_4(x)$ is irreducible over $\mathbb{Q}$ 
and $\sharp H$ takes the maximum value $\sharp H = N_m = 12$. 
In this case we obtain the Galois group 
$G = \mathbb{Z}_2 \ltimes \, (\mathbb{Z}_2 
                   \times \mathbb{Z}_2 \times \mathbb{Z}_3)$. 
Their corresponding generators $b$, $a_1$, $a_2$ and 
$a_3$ hold the relations 
\begin{equation}
b \, a_1 \, b = a_2, \qquad 
              b \, a_3 \, b = a_3^{-1} 
\end{equation}
and $a_1$, $a_2$ and $a_3$ are commutable 
with each other.

The Galois group $G$ is, as described in the next section, 
the extension of $\mathbb{Z}_2$ by an abelian group $H$, 
$G/H \cong \mathbb{Z}_2$. 
There are many types of the Galois group $G$. 
As seen in explicit examples, there exist both the abelian 
and the nonabelian $G$. 
In many nonabelian cases, $G$ contains the dihedral 
group $D_n \ (n \geq 3)$ but not the symmetric group 
$S_n \ (n \geq 4)$. 
In the next section and Appendix we discuss the detailed 
classification of the Galois group $G$ given by 
the extension of $\mathbb{Z}_2$ by an abelian group $H$.

\vspace{5mm}

\section{Extension of $\mathbb{Z}_2$ by an abelian kernel $H$}

In the above we have proposed a hypothesis 
that the Galois group $G=Gal(L/\mathbb{Q})$ of the Galois 
extension field $L=K(E_{\Lambda}[\mathfrak{a}])$, 
generated by adjoining the torsion points $E_{\Lambda}[\mathfrak{a}]$ 
of the elliptic curve $E_{\Lambda}$ to the quadratic imaginary 
field $K = \mathbb{Q} (\tau)$, 
should be regarded as a flavor symmetry in the effective 
theory from string. 
The group $G$ is settled according as the choice of an integral 
ideal $\mathfrak{a} $ of $ R_K$. 
In the previous section we have taken $\mathfrak{a} = m R_K$ and 
$E_{\Lambda}[\mathfrak{a}]$ to be the $m$-torsion points. 
Moreover, we have calculated the group $G$ explicitly 
in the cases of $m=3,4$. 
It is, however, difficult to implement the same calculation 
for larger values of $m$. 
In this section we investigate general properties of the Galois 
group $G$ of the extension $L/\mathbb{Q}$. 
We will see that the Galois group $G=Gal(L/\mathbb{Q})$ is 
an extension of $ \mathbb{Z}_2$ by some abelian group $H$ 
and that when $\mathfrak{a} = m R_K$, this extension is 
a semidirect product group $\mathbb{Z}_2 \ltimes H$.

\subsection{Fundamental theorem of Galois theory}
The Galois group $G=Gal(L/\mathbb{Q})$ has an important property 
described by the fundamental theorem of Galois theory.\cite{Lang} 
This theorem states that, in a Galois extension $L/F$, 
any intermediate extension $K/F$ is also a Galois extension 
if and only if the group $Gal(L/K)$ is a normal subgroup of 
$Gal(L/F)$. 
The theorem also asserts that the three groups are related 
to each other as $Gal(L/F)/Gal(L/K)\cong Gal(K/F)$. 
Applying the theorem to the three number fields 
$L=K(E_{\Lambda}[\mathfrak{a}])$, $K=\mathbb{Q}(\tau)$ 
and $F=\mathbb{Q}$, 
we obtain the isomorphism 
$G/H\cong Gal(K/\mathbb{Q}) =\mathbb{Z}_2$. 
As stated in section 3, the group $H=Gal(L/K)$ is abelian.
Thus, $G=Gal(L/\mathbb{Q})$ is an extension of $ \mathbb{Z}_2$ 
by the abelian kernel  $H=Gal(L/\mathbb{Q}(\tau))$.

We further notice that the field $L=K(E_{\Lambda}[\mathfrak{a}])$ 
contains the field $K=\mathbb{Q}(\tau)$ and that the group 
$Gal(K/\mathbb{Q})$ is generated by complex conjugation map. 
This map forms a $\mathbb{Q}$-automorphism of $L$ 
if $L$ is obtained by adjoining the roots of an algebraic equation 
with coefficients of rational numbers to $K$. 
In such case $\mathbb{Z}_2 \cong Gal(K/\mathbb{Q}) $ is 
a subgroup of $Gal(L/\mathbb{Q})$. 
When $\mathfrak{a} = m R_K$, this is the case. 
For a general integral ideal $\mathfrak{a}$, however, 
$\mathbb{Z}_2 $ is not necessarily a subgroup of $G$.

\subsection{Semidirect product}
In what follows, we focus on the case that $ \mathbb{Z}_2$ is 
 homomorphically embedded in $G$. 
It is well known in group theory that, in the extension $G$ of 
a group $Q$ by a group $H$, $G/H\cong Q$, 
$G$ is a semidirect product $G = Q\ltimes_{\sigma} H$ 
if and only if $Q$ can be  homomorphically embedded 
in $G$.\cite{Rotman}
The subscript $\sigma$ refers to a homomorphism from $Q$ to 
the automorphism group of $H$, $Q \to \mathrm{Aut}\,H$ 
denoted by $b \mapsto \sigma_{b}$. 
The semidirect product  $ Q\ltimes_{\sigma} H$ is described 
in terms of Cartesian product $\{(b,h) \}$ in which 
the product of two elements $(b,h)$ and $(b',h')$ is defined as
\begin{equation}
(b,h) \cdot (b',h') = (b\,b',\sigma_{b'}(h)\,h')  
\label{eq:semidirect} 
\end{equation}
with $\sigma_{b'}(h) = b' \, h \, {b'}^{-1}$. 
Identifying $b\in Q$ and $h \in H$ with 
$(b,e), \, (e,h) \in Q \times H$ respectively, 
we obtain 
\begin{gather}
b^{2} = (b,e)\cdot(b,e) = (b^{2},\sigma _{b}(e) e) = (b^{2},e), 
                                                 \label{eq:f1} \\
b\,h\,b = (b,e)\cdot(e,h)\cdot(b,e) = (b,h)\cdot(b,e) 
                         = (b^{2},\sigma_{b}(h)). \label{eq:f2} 
\end{gather}
In our setting $Q=\mathbb{Z}_2$, the homomorphism $\sigma$ 
satisfies $\sigma\cdot\sigma=\mathrm{id}_{\mathbb{Z}_2}$. 
Because the nontrivial element of $\mathbb{Z}_2$ is only its 
generator, we shall hereafter denote the generator simply as $b$. 
Then Eqs.~\eqref{eq:f1} and  \eqref{eq:f2} become 
\begin{equation}
{b}^{2} = e,\quad {b}\,h\,b = \sigma_{b}(h). 
\label{eq:fundamental} 
\end{equation}
Equation \eqref{eq:fundamental} constitutes fundamental 
relations of the group $\mathbb{Z}_2\ltimes_{\sigma} H$.

\subsection{Examples}
The fundamental theorem of abelian groups states that 
a finite abelian group $H$ is, in general, isomorphic to 
a product of cyclic groups of the form
\[
  H \cong \mathbb{Z}_{N_1} \times 
          \mathbb{Z}_{N_2} \times  \dotsm \times 
          \mathbb{Z}_{N_t},  
\]
where $N_{i} = {p_i}^{e_i} \, (p_i \leq p_{i +1} )$, 
$e_i \in \mathbb{Z}_{+}$ and $p_i $'s are prime numbers. 
In the remaining part of this section, we explore the semidirect 
product for the cases of $t=1,2$. 
Since the semidirect product  $\mathbb{Z}_2 \ltimes_{\sigma} H$ 
is determined by $\sigma_{b} \in \mathrm{Aut}\,H$ 
as dictated in Eq.\eqref{eq:fundamental}, 
we examine the possible automorphisms of order two.

\subsubsection{$H=\mathbb{Z}_{N}  \ \ (N=p^{e}) $}

An automorphism $\sigma_{b} $ of $H$ is given by 
$\sigma_{b} (a) = a^{n}$ for  a generator $a$ of $\mathbb{Z}_{N}$. 
The relation  $\sigma_{b}\cdot\sigma_{b}=\mathrm{id}_{\mathbb{Z}_2}$ 
implies $(a^{n})^{n}=a^{n^{2}}= a$, whose solutions are 
$n_{0} \equiv \pm 1,  \ \pm r \pmod{N} $, 
where $r = 2^{e - 1} + 1$. 
The latter solution $n_{0} \equiv \pm r$ are possible only 
when $p=2$ and $e \geq 3$. 
The semidirect product $G$ becomes 
\[
G  =\bigl \langle a,b \bigm | a^{N} = b^{2}= e, \; 
                      b ab = a^{n_{0} }  \bigr \rangle . 
\]
The group $G$ with $n_{0} \equiv 1$ is the direct product group 
and  $G$ with $n_{0} \equiv -1$ is the dihedral group.

\subsubsection{$H=\mathbb{Z}_{N_1} \times\mathbb{Z}_{N_2} \ \ 
          (N_1={p_1}^{e_1}, \ N_2={p_2}^{e_2}, \ e_1 \leq e_2) $}

An automorphism  $\sigma_{b} $ is given by 
\begin{equation}
\left\{\begin{array}{c}
\sigma_{b} (a_1) = a_{1}^{m_1}\,a_{2}^{m_2}, \\
\sigma_{b} (a_2) = a_{1}^{n_1}\,a_{2}^{n_2}, 
\end{array}\right.       
\label{eq:sigma} 
\end{equation}
where $a_1$ and $a_2$ are generators of $\mathbb{Z}_{N_1}$ and 
$\mathbb{Z}_{N_2}$, respectively. 
In exploring all the possible automorphism of order two, 
we should pay our attention to the following four points. 

\begin{itemize}
\item \textit{well-definedness of }$\sigma_{b}$ \\
We note that $m_1,n_1$ are defined in$\mod{N_1}$ and $m_2,n_2$ 
in$\mod{N_2}$, respectively.
Automorphic property of $\sigma_{b} $ leads to 
\begin{eqnarray*}
e = \sigma (a_{1}^{N_1}) = \sigma (a_1)^{N_1} = a_{2}^{m_2 N_1}, \\ 
e = \sigma (a_{2}^{N_2}) = \sigma (a_2)^{N_2} = a_{1}^{n_1  N_2} ,
\end{eqnarray*}
which requires 
\begin{eqnarray*}
m_2 N_1  \equiv 0 \ \ \pmod{N_2}, \\
n_1 N_2  \equiv 0  \ \  \pmod{N_1}.
\end{eqnarray*}
From these congruence equations, we find that 
$\text{if} \;  p_1 \neq p_2$, then 
\begin{equation}
 n_1 \equiv 0 \pmod{N_1},\quad m_2 \equiv 0  \pmod{N_2}
\label{eq:mod2}, 
\end{equation}
and that, $\text{if} \;  p_1 = p_2 =: p$, then
\begin{equation}
p^{e_2 -e_1}| m_2 \label{eq:mod2}.
\end{equation}

\item \textit{bijectivity of}  $\sigma_{b} $ \\
The case $ p_1 \neq p_2$ is essentially reduced to the case 
given in subsection 4.3.1. 
The case $ p_1 = p_2 =: p$ requires some scrutiny.
The homomorphism  $\sigma_{b} $ given in \eqref{eq:sigma} 
is bijective only when we have the relations 
\begin{equation*}
\left\{\begin{array}{c} 
a_1 = \sigma_{b} (a_{1}^{s_1}\,a_{2}^{s_2}), \\
a_2 = \sigma_{b} (a_{1}^{t_1}\,a_{2}^{t_2})
\end{array}\right.
\end{equation*}
for some $s_1,s_2,t_1,t_2 \in\mathbb{Z}$.
These relations are translated into 
\begin{equation}
\left\{
\begin{array}{cl}
m_1 s_1 + n_1 s_2 \equiv 1 & \ \ \pmod{N_1}, \\
m_2 s_1 + n_2 s_2 \equiv 0 & \ \ \pmod{N_2}, \\
m_1 t_1 + n_1 t_2 \equiv 0 & \ \ \pmod{N_1}, \\
m_2 t_1 + n_2 t_2 \equiv 1 & \ \ \pmod{N_2}. 
\end{array}
\right. 
\label{eq:bijectivity}
\end{equation}
From these equations we can easily find that 
\[
\sigma_{b} \ \text{is bijective} \iff 
              m_1 n_2 - m_2 n_1 \not\equiv 0 \pmod{p}. 
\]

\item 
$\sigma_{b}\cdot\sigma_{b}=\mathrm{id}_{\mathbb{Z}_2}$ \\
This leads to 
\begin{equation}
\left\{
\begin{array}{cl}
{m_1}^2 + n_1m_2 \equiv 1 & \ \ \pmod{N_1}, \\
m_2 (m_1 + n_2 ) \equiv 0 & \ \ \pmod{N_2}, \\
n_1 (m_1 + n_2 ) \equiv 0 & \ \ \pmod{N_1}, \\
n_1m_2 + {n_2}^2 \equiv 1 & \ \ \pmod{N_2}. 
\end{array}
\right.\label{eq:ID}
\end{equation}

\item \textit{change of generators} \\
The matrix 
$A:=\bigl( \begin{smallmatrix} m_1 &  n_1 \\ 
                m_2 & n_2\end{smallmatrix} \bigr)$
which describes the automorphism $\sigma_{b}$ depends on 
the choice of generators of $H$. 
If we change the generators of $H$ from $( a_{1} , a_{2} )$ 
to $( a_{1}' , a_{2}' )$ as
\begin{equation*}
\left\{\begin{array}{c} 
 a_{1}  = {a_{1}'}^{\alpha}\, {a_{2}'}^{\gamma}, \\
 a_{2}  = {a_{1}'}^{\beta}\, {a_{2}'}^{\delta}, 
 \end{array}\right.
\end{equation*}
then the matrix $A$ changes 
as\footnote{A similar reasoning which leads to Eq.~\eqref{eq:mod2} 
requires that $p^{e_{2}-e_{1}}$ divides $\gamma$.}
\[
A \to A'= V A V^{-1}, \quad 
V=\bigl( \begin{smallmatrix} \alpha & \beta \\ 
                \gamma & \delta \end{smallmatrix} \bigr). 
\]

Hence we should regard $A$ and its conjugate $V A V^{-1}$ as 
equivalent to each other. 
\end{itemize}

In the following we give some examples of the representative 
systems of the conjugacy classes and the corresponding 
semidirect product group $G$. 

\begin{enumerate}
\item {$p_1 < p_2$}
\[
A = \left(
\begin{array}{cc}
m_{1} & 0 \\
0 & n_{2}
\end{array}
\right),
\]
where 
\begin{align*}
m_{1} &=\left\{\begin{array}{l}
\pm 1\\ 
2^{e_1 - 1} \pm1 \quad (e_1 \geq 3,\ p_1 =2),
\end{array}
\right. \\
n_{2}  &= \pm 1.
\end{align*}
The semidirect product group is
\begin{equation*}
G  =\bigl \langle a_{1},a_{2},b \bigm | 
     {a_{1}}^{N_{1}} = {a_{2}}^{N_{2}}  =  b^{2}= e, \; 
     b \, a_{1} \,b = {a_{1}}^{m_{1}},  \; 
     b \, a_{2} \,b = {a_{2}}^{n_{2}} \bigr \rangle . 
\end{equation*}

\item{$p_1 = p_2 = p \neq 2$} \\
The set of  matrices $A$ falls into four conjugacy classes and 
their representatives can be chosen as 
\[
\left(\begin{array}{cc}
1 & 0 \\ 
0 & 1
\end{array}\right)\ , \ \ 
- \left(\begin{array}{cc}
1 & 0 \\
0 & 1
\end{array}\right)\ ,\;
\left(\begin{array}{cc}
1 & 0 \\
0 & -1
\end{array}\right)\ , \ \  
-\left(\begin{array}{cc}
1 & 0 \\
0 & -1
\end{array}\right)\ .
\]
and the semidirect product groups are
\begin{equation*}
G  =\bigl \langle a_{1},a_{2},b \bigm | 
{a_{1}}^{N_{1}} = {a_{2}}^{N_{2}}  =  b^{2}= e, \; 
b\, a_{1} \,b = {a_{1}}^{\pm 1},  \; 
b\, a_{2} \,b = {a_{2}}^{\pm1} \bigr \rangle .  
\end{equation*}

\item{$p_1 = p_2 =p = 2$}
\begin{enumerate}
\item {$N_1 = N_2 = N = 2^{e}$}
\begin{itemize} 
\item The case $|A| \equiv 1  \pmod{N}$
 
We have eight conjugacy classes and a representative system is 
\[
\begin{array}{cccc}
\pm \left(
 \begin{array}{cc}
  1 & 0 \\
  0 & 1
 \end{array}\right),\;
\pm r \left(
 \begin{array}{cc}
 1 & 0 \\
 0 & 1
 \end{array}\right), \;
\pm \left(
 \begin{array}{cc}
 1 & 2^{e - 1} \\
 0 & 1
 \end{array}\right),\;
\pm \left(
 \begin{array}{cc}
 1 & 2^{e - 1} \\
 2^{e - 1} & 1 
 \end{array}\right).
\end{array}
\]
\item  The case $|A| \equiv r \  \pmod{N}$ \\
We have four conjugacy classes and a representative system is 
\[
\begin{array}{cc}
\pm \left(
 \begin{array}{cc}
 r  & 0 \\
 0 & 1
 \end{array}\right),\;
\pm \left(
 \begin{array}{cc}
 1 & 2^{e - 1} \\ 
 2^{e - 1} & r 
 \end{array}\right). 
\end{array}
\]

\item The case $|A| \equiv -1 \  \pmod{N}$ \\
The matrix $A$ is of the form
\begin{eqnarray*}
A = \left(\begin{array}{cc}a & b \\ c & - a\end{array}\right) \\
\end{eqnarray*}
with $a^{2} + b c \equiv 1  \pmod{N}$. 
It is difficult to carry out complete classification. 
We only point out that there are at least the following three 
inequivalent classes; 
\begin{align*}
\left(
 \begin{array}{cc}
 1 & 0 \\
 0 & - 1 
 \end{array}\right) & \sim
- \left(
 \begin{array}{cc}
 1 & 0 \\
 0 & -1
 \end{array}\right), \\
\left(
 \begin{array}{cc}
 0 & 1 \\
 1 & 0
 \end{array}\right) & \sim
- \left(
 \begin{array}{cc}
 0 & 1 \\
 1 & 0
 \end{array}\right), \\
r \left(
 \begin{array}{cc}
 1 & 0 \\
 0 & - 1 
 \end{array}\right) &\sim
- r \left(
 \begin{array}{cc}
 1 & 0 \\
 0 & - 1 
 \end{array}\right). 
\end{align*}

\item The case $|A| \equiv - r \ \pmod{N}$ \\
Some matrices inequivalent to each other are 
\[
\pm \left(
 \begin{array}{cc}
 1 & 0 \\
 0 & - r 
 \end{array}\right). 
\]
\end{itemize}

\item{$N_1 = 2^{e_1}, \; N_2 = 2^{e_2} \quad ( e_1<  e_2)$} \\
There exist matrices $A$ which can not be diagonalized by some 
regular matrix $V$. 
An example of such type of the matrix is 
\[
  A=\left(
  \begin{array}{cc}
   1 & 1 \\
   0 & -1
  \end{array}\right)
\]
for the cases $(e_1,e_2) = (1,2),\, (2,3)$. 
\end{enumerate}
\end{enumerate}

\vspace{5mm}

\section{A phenomenologically viable example}

In this section we consider the effective theory in which 
the quark/lepton superfields and the Higgs superfields 
live in $M_4 \times T^2$, 
where $T^2$ is given by an elliptic curve $E_{\Lambda}$. 
It is postulated that $M_4 \times T^2$ is 
the intersection of two kinds of D-brane. 
For instance, we take the D-brane configuration in which 
one of the D-branes is $M_4 \times Y_4$ and the other is 
$M_4 \times Y'_4$, where $Y_4$ and $Y'_4$ are different 
4-dimensional extra spaces with $Y_4 \cap Y'_4 = T^2$. 
In this paper it is assumed that the extra space is discrete and 
that the extra space is all of the $m$-torsion points 
$E_{\Lambda}[m]$ of the elliptic curves $E_{\Lambda}$ with CM. 
Our point of view is that the Galois group is nothing but 
the flavor symmetry for open strings. 
Thus, we explore the Galois group $G$ relevant 
to the underlying flavor symmetry including the R-parity. 
The flavor symmetry yields the selection rule for interactions 
among quarks and leptons and then governs the stability of proton 
and the characteristic texture of fermion masses and mixings.

When we take some elliptic curve $E_{\Lambda}$ with CM and some 
number $m$ which represents the size of $E_{\Lambda}$ in unit 
of $E_{\Lambda_0}$, 
the Galois extension field $L = K(E_{\Lambda}[m])$ 
induces the abelian Galois group $H = Gal (L/K)$ and also 
the Galois group $G = Gal (L/\mathbb{Q})$. 
The latter is an extension of $\mathbb{Z}_2 = Gal (K/\mathbb{Q})$ 
by $H$ and is a semidirect product group $\mathbb{Z}_2 \ltimes H$. 
We now assume that there exists an elliptic curve $E_{\Lambda}$ 
with CM such that the following groups $H$ and $G$ are 
the Galois groups $Gal (L/K)$ and $Gal (L/\mathbb{Q})$ 
derived from the Galois extension $L = K(E_{\Lambda}[m])$: 
\begin{eqnarray}
 & & H = \mathbb{Z}_4 \times \mathbb{Z}'_4 \times \mathbb{Z}_N, \\
 & & G = \{ \mathbb{Z}_2 \ltimes \, (\mathbb{Z}_4 
             \times \mathbb{Z}'_4) \} \times \mathbb{Z}_N. 
\end{eqnarray}
Here $N$ is taken as odd. 
Let us denote the generators of $\mathbb{Z}_2$, $\mathbb{Z}_4$, 
$\mathbb{Z}'_4$ and $\mathbb{Z}_N$ by $b$, $a_1$, $a_2$ and 
$a_3$, respectively. 
These generators hold the relations 
\begin{equation}
b \, a_i \, b = a_i^{-1} \ \ (i = 1, \ 2), 
     \qquad b \, a_3 \, b = a_3 
\end{equation}
and $a_1$, $a_2$ and $a_3$ are commutable 
with each other. 
Furthermore, we suppose that one of two kinds of the D-brane has 
the degree of freedom of $SU(6)$ gauge group and 
the other has that of $SU(2)_R$ gauge group. 
Therefore, we study the $SU(6) \times SU(2)_R$ model with 
the flavor symmetry designated by the Galois group $G$.

Here we briefly summarize the parts of 
the $SU(6) \times SU(2)_R$ string-inspired model which 
are relevant to our present study. 
For a more detailed description see Refs.\cite{Matsu}. 

\begin{enumerate}
\item The gauge group $SU(6) \times SU(2)_R$ can be 
obtained from $E_6$ through the $\mathbb{Z}_2$ flux breaking 
on a multiply-connected manifold. 
In contrast to the conventional GUT-type models, 
we have no Higgs fields of adjoint or higher representations. 
Nevertheless, the symmetry breaking of $SU(6) \times SU(2)_R$ 
down to the standard model gauge group can take place 
via the Higgs mechanism.

\item It is assumed that the matter content consists of 
the chiral superfields of three families and 
the single vector-like multiplet in the form 
\begin{equation}
  3 \times {\bf 27}(\Phi_{1,2,3}) + 
        ({\bf 27}(\Phi_0)+\overline{\bf 27}({\bar \Phi})) \nonumber 
\end{equation}
in terms of $E_6$. 
The superfields $\Phi$ in {\bf 27} of $E_6$ are decomposed into 
two irreducible representations of $SU(6) \times SU(2)_R$ as 
\begin{equation}
  \Phi({\bf 27})=\left\{
       \begin{array}{lll}
         \phi({\bf 15},{\bf 1})& : 
               & \quad \mbox{$Q,L,g,g^c,S$}, \\
          \psi({\bf 6}^*,{\bf 2}) & : 
               & \quad \mbox{$(U^c,D^c),(N^c,E^c),(H_u,H_d)$}, 
       \end{array}
       \right. \nonumber 
\end{equation}
where the pair $g$ and $g^c$ and the pair $H_u$ and $H_d$ represent 
the colored Higgs and the doublet Higgs superfields, respectively. 
$N^c$ is the R-handed neutrino superfield and 
$S$ is an $SO(10)$ singlet. 
Note that the doublet Higgs and the color-triplet 
Higgs fields belong to different irreducible representations 
of $SU(6) \times SU(2)_R$ and so the triplet-doublet splitting 
problem is solved naturally.

\item There are only two types of gauge invariant trilinear 
combinations 
\begin{eqnarray}
   (\phi ({\bf 15},{\bf 1}))^3 & = & QQg + Qg^cL + g^cgS, \nonumber \\
    \phi ({\bf 15},{\bf 1})(\psi ({\bf 6}^*,{\bf 2}))^2 & 
               = & QH_dD^c + QH_uU^c + LH_dE^c            \nonumber \\ 
             {}& & \qquad  + LH_uN^c + SH_uH_d + gN^cD^c  \nonumber \\ 
             {}& & \qquad  + gE^cU^c + g^cU^cD^c.  \nonumber 
\end{eqnarray}
\end{enumerate}

We now address ourselves to a discussion of the flavor-charge 
assignment to each matter superfield. 
To begin with, we argue the relation between 
$Gal (K/\mathbb{Q}) = \mathbb{Z}_2$ and the R-parity. 
The breaking scale of the $SU(6) \times SU(2)_R$ gauge 
symmetry should be larger than ${\cal O}(10^{16}{\rm GeV})$ 
to guarantee the longevity of the proton. 
Here, it is worth recalling the D-flat condition. 
When the gauge symmetry is broken at a large scale, 
the D-flat condition allows the supersymmetry to remain 
unbroken down to a TeV scale. 
In order to guarantee the D-flatness with leaving the R-parity 
unbroken, the effective theory has to contain conjugate pairs 
of R-parity even matter superfields relative to the gauge group. 
It is the pair of R-parity even matter superfields 
that acquires VEV along a D-flat direction. 
As a consequence of the gauge symmetry breaking at the 
large scales, most of the R-parity even matter superfields 
lie at the energy scales much larger than ${\cal O}(1 {\rm TeV})$. 
By contrast, among R-parity odd matter superfields 
with three generations the particles of MSSM have to 
remain massless as low as ${\cal O}(1 {\rm TeV})$. 
Then, there should be no conjugate pairs of the R-parity odd 
matter superfields. 
This means that the R-parity is the quantum number which 
discriminates between matter superfields with conjugate pair 
and those without pair. 
Here we notice that the interchange of the generation matter 
superfields with the anti-generation ones corresponds to 
the complex conjugation transformation for Calabi-Yau manifold. 
On the other hand, in view of the fact that $K$ is a quadratic 
imaginary field, the automorphism with conjugate properties 
on the elliptic curves with CM is the subgroup 
$Gal (K/\mathbb{Q}) = \mathbb{Z}_2$ of the Galois group $G$. 
Therefore, we identify $Gal (K/\mathbb{Q}) = \mathbb{Z}_2$ 
with the R-parity.

Next we turn to a discussion of the charge assignment for 
$Gal (L/K) = H = \mathbb{Z}_4 \times \mathbb{Z}'_4 \times \mathbb{Z}_N$. 
We rewrite $H$ as 
\begin{equation}
  H \cong \mathbb{Z}_4 \times \mathbb{Z}_{4N} 
     = \langle a_1 \rangle \times \langle \widetilde{a}_3 \rangle 
\end{equation}
and denote the charge of each matter superfield by 
the notations presented in Table 5. 
Note that $N =$ odd, so $2N+1 \equiv -1 \ \pmod{4}$. 
Consequently, the generator $\widetilde{a}_3$ of 
$\mathbb{Z}_{4N}$ holds the relation 
\begin{equation}
  b \, \widetilde{a}_3 \, b = \widetilde{a}_3^{\ 2N+1}. 
\end{equation}
Here we assign flavor-charges to matter superfields 
so as to obtain phenomenologically viable solutions. 
As in existing phenomenological models, in our approach 
we cannot definitely settle flavor-charges of 
matter fields from theoretical arguments. 
However, we require the mixed-anomaly conditions. 
These conditions are stringent and the solutions are 
rather restricted. 
The flavor-charge of the Grassmann number $\theta$ is assigned as 
$b \, a_1 \, \widetilde{a}_3^{\ q_{\theta}}$. 
In general, the couplings among matter superfields are 
given by the nonrenormalizable terms suppressed by powers of 
${\cal O}(M_S^{-1})$. 
The explicit forms of the nonrenormalizable terms are 
determined by the flavor symmetry.

\begin{table}[t]
\caption{Flavor-charge assignment to matter superfields.}
\label{table:5}
\begin{center}
\begin{tabular}{|c|c|cc|} \hline 
    & \phantom{MM} $\Phi_i \ (i=1,2,3)\ $ & 
        \phantom{MM} $\Phi_0$ \phantom{MM} & 
              \phantom{MM} $\bar{\Phi}$ \phantom{MM} \\ \hline
$\phi({\bf 15, \ 1})$  & $b \, \widetilde{a}_3^{\ q_i}$  & 
            $\widetilde{a}_3^{\ q_0}$  &  $\widetilde{a}_3^{\ \bar{q}}$ \\
$\psi({\bf 6^*, \ 2})$ & $b \, a_1^2 \, \widetilde{a}_3^{\ r_i}$  & 
    $a_1^2 \, \widetilde{a}_3^{\ r_0}$  & 
          $a_1 \, \widetilde{a}_3^{\ \bar{r}}$ \\ \hline
\end{tabular} 
\end{center}
\end{table}

Under the charge assignment given in Table 5, 
the superpotential in the R-parity even sector can be settled as 
\begin{equation}
  W_1 = M_S^3 \left[ c_0 
     \left( \frac{\phi_0 \bar{\phi}}{M_1^2} \right)^{2n} 
       + c_1 \left( \frac{\phi_0 \bar{\phi}}{M_1^2} \right)^n 
         \left( \frac{\psi_0 \bar{\psi}}{M_2^2} \right)^4 
       + c_2 \left( \frac{\psi_0 \bar{\psi}}{M_2^2} \right)^8 \right], 
\end{equation}
where $M_S, \ M_1, \ M_2 = {\cal O}(10^{19}{\rm GeV})$ represent 
the string scale and $c_i = {\cal O}(1)$. 
Note that $\phi_0 \, \bar{\phi}$ and $\psi_0 \bar{\psi}$ 
have both nonzero flavor-charges. 
As seen from Table 5, the exponent $2n \ (> 0)$ of the first 
term on the r.h.s. is determined through 
the $\mathbb{Z}_{4N}$-invariance. 
The exponent of the third term on the r.h.s. is constrained 
both by the $\mathbb{Z}_4$-invariance and 
by the $\mathbb{Z}_{4N}$-invariance. 
We choose $2n \sim 4N \gg 8$. 
Then, we carry out the minimization of the scalar potential 
with the soft SUSY breaking mass terms characterized by the 
scale ${\cal O}(1{\rm TeV})$. 
The $D$-flat solution implies\cite{Dflat} 
\begin{equation}
   |\langle \phi_0 \rangle| = |\langle \bar{\phi} \rangle| 
    > |\langle \psi_0 \rangle| = |\langle \bar{\psi} \rangle|. 
\end{equation}
Introducing the notation 
\begin{equation}
   \xi = \frac{| \langle \phi_0 \rangle \, 
              \langle \bar{\phi} \rangle |}{M_1^2}, \nonumber 
\end{equation}
we have 
\begin{equation}
   \xi^{2n} = {\cal O}(10^{-17}) = {\cal O}(\frac{m_W}{M_S}), \qquad 
   \frac {\langle \psi_0 \rangle \langle {\overline \psi} \rangle}{M_1^2} 
           = \xi^{\frac{n}{4} + \delta_N}   \nonumber 
\end{equation}
with $\xi^{\delta_N} = {\cal O}(1)$. 
Consequently, the spontaneous breaking of the gauge symmetry 
occurs in two steps as 
\begin{equation}
   SU(6) \times SU(2)_{\rm R} 
     \buildrel \langle \phi_0 \rangle \over \longrightarrow 
             SU(4)_{\rm PS} \times SU(2)_L \times SU(2)_{\rm R}  
     \buildrel \langle \psi_0 \rangle \over \longrightarrow 
     G_{\rm SM}, 
\end{equation}
where $SU(4)_{\rm PS}$ and $G_{SM}$ are the Pati-Salam $SU(4)$ 
group and the standard model gauge group, respectively. 
The VEVs turn out to be $|\langle \phi_0 \rangle | 
                    = {\cal O}(10^{18}{\rm GeV})$ and 
$|\langle \psi_0 \rangle | 
              = {\cal O}(10^{17}{\rm GeV})$.

The nonrenormalizable terms in the superpotential induce 
the low-energy effective interactions through the 
Froggatt-Nielsen mechanism.\cite{F-N} 
The superpotential terms which bring about the effective 
Yukawa couplings of $\Phi_i$ $(i = 0, \ 1, \ 2, \ 3)$ 
take the forms 
\begin{eqnarray}
  W_Y & = & \frac{1}{3!} \, z_0 \left( \frac{\phi_0 \bar{\phi}}
              {M_1^2} \right)^{\zeta_{00}} (\phi_0)^3 
          + \frac{1}{2} \, h_0 \left( \frac{\phi_0 \bar{\phi}}{M_1^2} 
               \right)^{\eta_{00}} \phi_0 \psi_0 \psi_0   \nonumber \\
     {}& & + \, \frac{1}{2} \, \sum_{i,j=1}^{3} z_{ij} 
                  \left( \frac{\phi_0 \bar{\phi}}{M_1^2} 
                  \right)^{\zeta_{ij}} \phi_0 \phi_i \phi_j   
     + \frac{1}{2} \, \sum_{i,j=1}^{3} h_{ij} \left( 
           \frac{\phi_0 \bar{\phi}}{M_1^2} \right)^{\eta_{ij}} 
                          \phi_0 \psi_i \psi_j    \nonumber \\
     {}& & \phantom{MMM} + \sum_{i,j=1}^{3} m_{ij} 
             \left( \frac{\phi_0 \bar{\phi}}
              {M_1^2} \right)^{\mu_{ij}} \psi_0 \phi_i \psi_j. 
\end{eqnarray}
The indices $i$ and $j$ run over the generation. 
Each coefficient in the r.h.s. is supposed to be ${\cal O}(1)$. 
Furthermore, the coefficients of the third, fourth and fifth terms 
are assumed to exhibit $3 \times 3$ matrices with rank 3. 
All of the exponents in the above equation are nonnegative 
integers and determined by the flavor symmetry. 
Although the parameter $\xi$ is not a very small number, 
the large hierarchy occurs by raising the $\xi$ to large powers. 
The colored Higgs mass is given by 
$z_0 \, \xi^{\zeta_{00}} \langle \phi_0 \rangle$. 
In order that the colored Higgs gains a large mass 
around ${\cal O}(10^{17}{\rm GeV})$, 
the exponent $\zeta_{00}$ should be sufficiently small 
compared with $N$. 
By contrast, the doublet Higgs mass of 
${\cal O}(10^2{\rm GeV})$ is given by 
$h_0 \, \xi^{\eta_{00}} \langle \phi_0 \rangle$. 
Then, the exponent $\eta_{00}$ should be nearly equal to 
$4N \sim 2n$. 
The hierarchical structure of fermion masses are derived 
via an appropriate flavor-charge assignment for 
each matter superfield. 
For instance, the mass matrix for up-type quarks comes from 
the effective Yukawa terms 
\begin{equation}
  {\cal M}_{ij} \, Q_i \, U_j^c \, H_{u0}, \qquad 
               {\cal M}_{ij} = m_{ij} \, \xi^{\mu_{ij}}. 
\end{equation}
In the down-type quark sector $D^c$-$g^c$ mixing occurs. 
Similarly, in the lepton sector there appears $L$-$H_d$ mixing. 
In both sectors there exist three heavy modes with their masses 
larger than ${\cal O}(10^{10}{\rm GeV})$ and three light modes 
appearing in the low-energy spectrum. 
In addition, in this model the seesaw mechanism is naturally 
at work. 
The Majorana masses of the R-handed neutrinos are induced 
from the nonrenormalizable term 
\begin{equation}
  W_M = \frac{f_N}{M_S} \, \left( \frac{\phi_0 \bar{\phi}}
                             {M_1^2} \right)^{\nu_{ij}} 
              (\psi_i \, \bar{\psi})(\psi_j \, \bar{\psi}) 
\end{equation}
with $f_N = {\cal O}(1)$.

In the above superpotential the exponent of each term 
is determined by the conditions of $\mathbb{Z}_{4N}$-invariance 
on the $F$-term 
\begin{equation}
\begin{array}{l}
2n \, d \equiv - 2 \, (N + 1) \, q_{\theta},  \\
8 \, (r_0 + \bar{r}) \equiv 2 \, (N + 1) \, q_{\theta},   \\
d \, \zeta_{00} \equiv 3q_0 - 2 \, (N + 1) \, q_{\theta},  \\
d \, \eta_{00}  \equiv q_0 + 2r_0 - 2 \, (N + 1) \, q_{\theta},  \\
d \, \zeta_{ij} \equiv (2 N + 1) \, q_i + q_j 
                        + q_0 - 2 \, (N + 1) \, q_{\theta},  \\
d \, \eta_{ij}  \equiv (2 N + 1) \, r_i + r_j 
                        + q_0 - 2 \, (N + 1) \, q_{\theta},  \\
d \, \mu_{ij}   \equiv (2 N + 1) \, q_i + r_j 
                        + r_0 - 2 \, (N + 1) \, q_{\theta},  \\
d \, \nu_{ij}   \equiv (2 N + 1) \, r_i + r_j + 2 \, (N + 1) \, \bar{r} 
           - 2 \, (N + 1) \, q_{\theta}, \quad \pmod{4N} 
\end{array}
\end{equation}
with $d = - (q_0 + \bar{q})$. 
Furthermore, the additional conditions should be satisfied 
as for the mixed anomalies $\mathbb{Z}_{M} \cdot (SU(6))^2$ and 
$\mathbb{Z}_{M} \cdot (SU(2)_R)^2$, 
where $M = 2, \ 4, \ 4, \ N$ corresponding to each subgroup of 
the Galois group $G = \{ \mathbb{Z}_2 \ltimes 
( \mathbb{Z}_4 \times \mathbb{Z}'_4 ) \} \times \mathbb{Z}_N$ in order. 
These conditions are expressed as\cite{Anom,Matsu2} 
\begin{equation}
  4 \, q_T + 2 \, r_T \equiv 18 \, q_{\theta}, \qquad 
  6 \, r_T \equiv 26 \, q_{\theta} \qquad \pmod{M}, 
\end{equation}
where 
\begin{equation}
q_T = \sum_{i=0}^{3} q_i + \bar{q}, \qquad 
r_T = \sum_{i=0}^{3} r_i + \bar{r}.   \nonumber 
\end{equation}

After some manipulation we find a solution with $N = 31$ 
which satisfies all of the above conditions and reproduces 
the characteristic texture of fermion masses and mixings. 
Another solution is found for the case $N = 35$. 
Here we concentrate on the case $N = 31$. 
By taking the conventional charge assignment of the 
Grassmann number $q_{\theta} = 1$, 
we obtain the $\mathbb{Z}_{4N}$-charge assignment to each 
matter superfield as shown in Table 6. 
Translating $\mathbb{Z}_{4N}$-charges into 
$\mathbb{Z}'_4 \times \mathbb{Z}_N$-charges, 
we present $\{ \mathbb{Z}_2 \ltimes ( \mathbb{Z}_4 \times 
\mathbb{Z}'_4 ) \} \times \mathbb{Z}_N$-charges in Table 7. 
The charge assignment of $q_{\theta}$ offered here implies 
that all of $\mathbb{Z}_2$, $\mathbb{Z}_4$, $\mathbb{Z}'_4$ and 
$\mathbb{Z}_{31}$ are the R-symmetry. 
In this solution the exponents of the effective interactions 
are given by $2n = 112$, $(\zeta_{00}, \ \eta_{00}) = (2, \ 116)$ 
and 
\begin{eqnarray}
  & &  \zeta_{ij} = \left(
       \begin{array}{ccc}
         34  &  30  &  22  \\
         30  &  26  &  18  \\
         22  &  18  &  10  
       \end{array}
       \right)_{ij},         \quad \ 
    \eta_{ij} = \left(
       \begin{array}{ccc}
         38  &  28  &  20  \\
         28  &  18  &  10  \\
         20  &  10  &   2  
       \end{array}
       \right)_{ij},      \nonumber \\
  & &  \mu_{ij} = \left(
       \begin{array}{ccc}
         31  &  21  &  13  \\
         27  &  17  &   9  \\
         19  &   9  &   1 
       \end{array}
       \right)_{ij},       \quad \ 
    \nu_{ij} = \eta_{ij} + 30. 
\end{eqnarray}
The $\xi$-parameter in this solution is related to 
the well-known $\lambda$ parameter in the CKM matrix 
as $\xi^4 \simeq \lambda = 0.22$. 
Taking $\delta_N = 3$, we have the fermion spectra\cite{Matsu} 
\begin{eqnarray}
  (m_u, \ m_c, \ m_t) & \simeq & (\xi^{31}, \ \xi^{17}, \ \xi ) 
                                                 \times v_u,  \\
  (m_d, \ m_s, \ m_b) & \simeq & (\xi^{31}, \ \xi^{22}, \ \xi^{10}) 
                                                 \times v_d, \\
  (m_e, \ m_{\mu}, \ m_{\tau}) & \simeq & 
                     (\xi^{31}, \ \xi^{15}, \ \xi^9) \times v_d, \\
  (m_{\nu_1}, \ m_{\nu_2}, \ m_{\nu_3}) & \simeq & 
            (\xi^8, \ \xi^6, \ 1 ) \times \frac{v_u^2}{M_S} \xi^{-31}, 
\end{eqnarray}
where $v_u$ and $v_d$ represent the VEVs $\langle H_{u0} \rangle$ 
and $\langle H_{d0} \rangle$, respectively. 
The neutrino mass for the third generation $m_{\nu_3}$ becomes 
${\cal O}(10^{-1}{\rm eV})$. 
Thus our model reproduces the orders of magnitude for fermion masses. 
The CKM matrix is of the form 
\begin{equation}
V_{\rm CKM} \simeq \left(
  \begin{array}{ccc}
                1      &   \lambda    &   \lambda^5  \\
            \lambda    &        1     &   \lambda^2  \\
            \lambda^3  &   \lambda^2  &    1 
  \end{array}
  \right). 
\end{equation}
Mixing angles in the MNS matrix are 
\begin{equation}
   \tan \theta_{12}^{\rm MNS} \simeq \xi,  \qquad 
     \tan \theta_{23}^{\rm MNS} \simeq \xi^3, \qquad 
       \tan \theta_{13}^{\rm MNS} \simeq \xi^4 \simeq \lambda. 
\end{equation}
This texture of the MNS matrix represents the LMA-MSW solution. 
All of these results are phenomenologically viable.

\begin{table}[t]
\caption{$\mathbb{Z}_{4N}$-charge \qquad $q_{\theta} = 1$}
\label{table:6}
\begin{center}
\begin{tabular}{|c|c|c|} \hline
\phantom{MMMM}  & \qquad $\phi({\bf 15}, \ {\bf 1})$ \qquad 
               & \qquad $\psi({\bf 6^*}, \ {\bf 2})$ \qquad \\ \hline 
$\Phi_0$      &   $q_0 = 18$      &  $r_0 = - 19$ \\
$\bar{\Phi}$  &   $\bar{q} = 49$  &  $\bar{r} = - 4$ \\ \hline 
$\Phi_1$      &   $q_1 = 62$      &  $r_1 = 52$ \\
$\Phi_2$      &   $q_2 = 82$      &  $r_2 = 102$ \\
$\Phi_3$      &   $q_3 = 122$     &  $r_3 = 18$ \\ \hline
\end{tabular}
\end{center}
\end{table}

\begin{table}[t]
\caption{$\{ \mathbb{Z}_2 \ltimes ( \mathbb{Z}_4 \times \mathbb{Z}'_4 ) \} 
\times \mathbb{Z}_N$-charge \qquad $q_{\theta} = (-, \ 1, \ 1, \ 1)$}
\label{table:7}
\begin{center}
\begin{tabular}{|c|c|c|} \hline
\phantom{MMMM}  & \phantom{MMM} $\phi({\bf 15}, \ {\bf 1})$ \phantom{MMM} 
       & \phantom{MMM} $\psi({\bf 6^*}, \ {\bf 2})$ \phantom{MMM} \\ \hline 
$\Phi_0$      &   $(+, \ 0, \ 2, \ 18)$  &  $(+, \ 2, \ 1, \ 12)$ \\
$\bar{\Phi}$  &   $(+, \ 0, \ 1, \ 18)$  &  $(+, \ 1, \ 0, \ 27)$ \\ \hline 
$\Phi_1$      &   $(-, \ 0, \ 2, \ \ 0)$ &  $(-, \ 2, \ 2, \ 21)$ \\
$\Phi_2$      &   $(-, \ 0, \ 2, \ 20)$  &  $(-, \ 2, \ 2, \ \ 9)$  \\
$\Phi_3$      &   $(-, \ 0, \ 2, \ 29)$  &  $(-, \ 2, \ 2, \ 18)$ \\ \hline
\end{tabular}
\end{center}
\end{table}

\vspace{5mm}

\section{Summary and discussion}

In this paper we have explored the underlying flavor symmetry 
through the arithmetic structure of elliptic curves with CM. 
There are close relations among RCFT, algebraic number theory 
and elliptic curves $E_{\Lambda}$ with CM. 
We have postulated that the extra space $E_{\Lambda}$, 
which is given by the intersection of two kinds of D-brane, 
is discretized and that the extra space is all of 
the $m$-torsion points $E_{\Lambda}[m]$ $(m \in \mathbb{Z}_+)$ 
of the elliptic curves $E_{\Lambda}$ with CM. 
The endpoints of open strings lie on $E_{\Lambda}[m]$. 
$E_{\Lambda}[m]$ provides the Galois 
extension field $L = K(E_{\Lambda}[m])$ which 
is the extension of field $K(j(E_{\Lambda}))$ generated by 
the coordinates of all of $E_{\Lambda}[m]$. 
Orientable open strings correspond to an automorphism of $L$. 
Hence we are led to the Galois group 
$G = Gal (L/\mathbb{Q}) = {\rm Aut}_{\mathbb{Q}}L$. 
Each element of $G$ acts on $E_{\Lambda}[m]$. 
Our point of view offered here is that $G$ is the flavor 
symmetry for open strings and that each element of $G$ 
corresponds to an orientable open string with free endpoints 
in the direction along the $E_{\Lambda}[m]$. 
We have studied the possible types of the Galois groups. 
It has been shown that the Galois group $G$ is an extension of 
$\mathbb{Z}_2$ by some abelian group $H$ and that $G$ is 
a semidirect product group $\mathbb{Z}_2 \ltimes H$ for 
the case of the $m$-torsion points of the elliptic curve. 
This is an important result of our approach. 
In general, $G$ is not an abelian group. 
$G$ possibly contains the dihedral group $D_n \ (n \geq 3)$ 
but not the symmetric group $S_n \ (n \geq 4)$. 
We have exhibited a phenomenologically viable example, 
in which the Galois groups $H = Gal (L/K)$ and $G = Gal (L/\mathbb{Q})$ 
are taken as $\mathbb{Z}_4 \times \mathbb{Z}'_4 \times \mathbb{Z}_{31}$, 
and $\{ \mathbb{Z}_2 \ltimes ( \mathbb{Z}_4 \times \mathbb{Z}'_4 ) \} 
\times \mathbb{Z}_{31}$, respectively. 
In the example it has been shown that the characteristic 
texture of fermion masses and mixings is reproduced 
and that the mixed-anomaly conditions are satisfied. 
In this paper we have assumed that there exists an elliptic curve 
$E_{\Lambda}$ with CM such that the Galois groups $H$ and $G$ 
offered here are derived from the $m$-torsion points of 
the elliptic curve. 

In the above viable example we have chosen 
$H = \mathbb{Z}_4 \times \mathbb{Z}'_4 \times \mathbb{Z}_N$ 
with $N = 31$. 
This case is possibly realized if we take $E_{\Lambda}[m]$ 
with $m = 311$. 
Note that $m = 311$ is a prime number, so $N_m = 311^2 - 1$. 
For the case $N = 35$ we have $m = 41$, 
in which $N_m = 41^2 - 1$. 
In both cases the relation $(\#H)|N_m$ is satisfied. 
The number $m$ represents the size of 
the extra space $E_{\Lambda}$ in unit of 
$E_{\Lambda_0}$ which corresponds to 
the fundamental scale of the string theory. 
The condition $m \leq {\cal O}(10^2)$ ensures that 
the extra space considered here is sufficiently small.

The Galois group $G = Gal(L/\mathbb{Q})$ 
is classified according as the homomorphism 
$\sigma : \ \mathbb{Z}_2 \rightarrow {\rm Aut} \, H$ 
and the factor set $f$. 
There are various types of the Galois group which 
we have not applied to a phenomenological model 
offered here. 
From phenomenological point of view, 
it is favorable for the flavor symmetry to contain 
the dihedral group $D_4$. 
This is due to the fact that the Majorana mass of 
the third generation is nearly equal to the geometrical 
average of $M_S$ and $M_Z$.\cite{Matsu} 
For the cases of the Galois group containing $D_4$ 
we explore the solutions in the same way as the present study. 
Under some phenomenological requirements we have to 
solve both conditions coming from the flavor symmetry 
and mixed-anomaly conditions. 
To do this, we need to use some ingenious manipulation. 
Further studies of phenomenologically viable Galois groups 
on elliptic curves with CM will be reported elsewhere. 
At present, when we find out the Galois group $G$ relevant to 
the flavor symmetry, the subsequent problem is open 
as to how to trace back to the Galois extension field $L$ and 
the elliptic curves $E_{\Lambda}$ with CM. 
However, our approach proposed here could shed new light on 
the study of the generation structure of quarks and leptons.

\vspace{5mm}


\appendix
\renewcommand{\thesection}{}
\renewcommand{\thesubsection}{A.\arabic{subsection}}
\section{Appendix \quad General theory of   group extension 
 and some examples}
\setcounter{equation}{0}
\renewcommand{\theequation}{A.\arabic{equation}}

In section 4 we have studied the group extension $G$ of 
$Q = \mathbb{Z}_2 $ by an abelian kernel $H$. 
In the study we have confined ourselves to the case 
when the quotient group $G/H \cong \mathbb{Z}_2$ is homomorphically 
embedded in $G$. 
In this Appendix we discuss the general case in which 
$\mathbb{Z}_2$ is not necessarily embedded as a subgroup in $G$.

\subsection{General theory of the extension 
               $H \vartriangleleft G \xrightarrow{\pi} Q$}
We sketch the general theory of an extension by an abelian kernel. 
For more detailed description see Ref.~\cite{de Azc}. 
We first start from the group $G$. 
Taking arbitrary section $s$ of the canonical projection 
$G \xrightarrow{\pi}Q$, 
we can write $g \in G$ as $g=hs(x)$ with $\pi(g) = x \in Q$, 
which yields the one-to-one correspondence 
$G \leftrightarrow Q \times H\; \text{by}\; g \mapsto (x,h)$. 
The homomorphic property of $\pi$ implies that $s(x)s(y)$ and $s(x y)$ 
are equal up to some element $f(x,y)$ of $\ker \pi  = H$, i.e.
\begin{equation}
s(x) s(y) = f(x,y) s(xy). \label{eq:factor set}
\end{equation}
The map $f:Q \times Q \to H$ is called the factor set of $G$ 
associated with the section $s$.
The product defined in $G$ induces a product in the set 
$Q \times H$ via the section $s$. 
In fact, from the equation
\begin{equation*}
g g' = h s(x) \, h' s(x') = h s(x) h' s(x)^{-1}s(x) s(x') 
                          =  h s(x) h' s(x)^{-1} f(x,x') s(xx') 
\end{equation*}
and $\pi\circ s = \mathrm{id}_{Q}$, we obtain the product 
\begin{equation*}
(x,h) (x', h') = ( xx',h s(x) h' s(x)^{-1} f(x,x') ). 
\end{equation*}
In view of the abelian nature of $H$, the automorphism 
$\sigma_{s(x)} := s(x)^{-1}\bullet s(x)$ does not depend on $s$. 
Hence, dropping  $s$, we can use the abbreviated notation 
\[
\sigma _{x} (h') = s(x)\, h' \, s(x)^{-1} =:  {^{x} h'}.
\]
Thus the product is rewritten as
\begin{equation}
 (x, \, h)(x', \, h') = (x \,x',\; h \, {^{x}h'} \; f(x,x')). 
\label{eq:product} 
\end{equation}
The associative law of  $G$ leads to the associative law of 
the above product in $H \times Q$, which requires 
the cocycle condition 
\begin{equation}
f(x,y) \,f(xy,z) = { ^{x} f(y,z)}\, f(x, yz). 
\label{eq:cocycle}
\end{equation}
If we change the section $s$ to $s'$
\[
s'(x) = \alpha (x) s(x)  \qquad \alpha (x)\in H,
\]
$f$ is transformed into 
\begin{equation}
f'(x,y) = \alpha (x)\,{ ^{x} \alpha (y)} \,\alpha (xy) ^{-1} f(x,y). 
\label{eq:section} 
\end{equation}
When this relation holds between $f$ and $f'$, 
we express as $f' \sim f$.

Reversing all the line of thought, i.e. starting from 
the set $Q \times H$, we can construct all of the group 
$G$ which give $G/H\cong Q$. 
If we introduce a homomorphism 
$\sigma : Q \to \mathrm{Aut}H,\quad x \mapsto 
  \sigma _{x} (\bullet) = { ^{x}(\bullet)}$ 
and a factor set $f:Q \times Q \to H$ which satisfies 
the cocycle condition~\eqref{eq:cocycle}, then we obtain 
the group $G^{\sigma}_{f}$ whose product is defined by 
\eqref{eq:product}. 
Note that if $f' \sim f$, $f'$ and $f$ give the same group, i.e.
\[
  G^{\sigma}_{f'} \cong G^{\sigma}_{f} \iff 
               f' (x,y) = \delta \alpha (x,y) f(x,y),\ \  
    \delta \alpha (x,y) := \alpha (x)\,{ ^{x} \alpha (y)} 
                                         \,\alpha (xy) ^{-1}. 
\]
The above argument proves the following theorem. 

\textbf{Theorem} \quad 
Let $Q$ be a group, $H$ be an abelian group and 
$\sigma$ be a homomorphism 
$\sigma : Q \to \mathrm{Aut}H,\quad x \mapsto { ^{x}(\bullet)}$, 
then the set of the group extensions of $Q$ by the kernel $H$ has 
the one-to-one correspondence with the equivalence class 
$H^{2}(Q,H) = Z^{2}(Q,H) / B^{2}(Q,H)$, where $Z^{2}(Q,H)$ and 
$B^{2}(Q,H)$ are the two-dimensional cocycle 
\[
 Z^{2}(Q,H) := \{f:Q \times Q \to H\, 
               |\, f(x,y) \,f(xy,z) = { ^{x} f(y,z)}\, f(x, yz) \}
\]
and the two-dimensional boundary 
\[
 B^{2}(Q,H) := \{ \delta \alpha : Q \times Q 
        \to H \, | \, \alpha : Q  \to H ; \delta \alpha (x,y) := 
            \alpha (x)\,{ ^{x} \alpha (y)} \,\alpha (xy) ^{-1}  \},
\]
respectively. \\ 
\textit{Remark.} If $f \sim e \in H$, we can choose the section $s$ 
such that $s(x)s(y) = s(xy)$. 
Using this homomorphism $s$, we can regard $Q$ as a subgroup of $G$. 
This means that the group $G^{\sigma}_{f}$ is a semidirect product. 
In section 4 we have studied only this case.

\subsection{Extensions of $Q = \mathbb{Z}_2$ by $H$}
Next we proceed to consider the case $Q = \mathbb{Z}_2$. 
Since $\mathbb{Z}_2 $ has only two elements $e$ and $ x_{1} $, 
the cocycle condition and the equivalence relation among 
the factor sets take a simple form. 
If $s(e) = e \in G$, then Eq.~\eqref{eq:factor set} leads 
to $f(x,e) = f(e,y) =e$. 
Nontrivial values of $f$ are possible only for $f(x_{1},x_{1})$, 
which is denoted simply by $f$. 
A change of the section $s(x) \to s'(x) = \alpha (x) s(x)$ 
is allowed only for $\alpha (x_{1}) $, which is denoted by $\alpha$. 
The cocycle condition~\eqref{eq:cocycle} becomes 
\begin{equation}
f = \sigma_{x_{1}} (f) 
\label{eq:cocycle2} 
\end{equation}
and the allowed change of the section induces the change of $f$ as 
\begin{equation*}
 f \to f' = \alpha \,\sigma_{x_{1}}  (\alpha) \, f, 
\end{equation*}
which means the equivalence relation 
\begin{equation}
 f \sim \alpha \,\sigma (\alpha)_{x_{1}}  \, f .
\label{eq:equiv}
\end{equation}
The product given by Eq.~\eqref{eq:product} leads to the relation 
among $b:= s(x_1)$ and $h \in H$ as 
\begin{equation}
{b}^{2} = f, \quad b \,h \,b^{-1} = \sigma_{x_{1}}  (h). 
\label{eq:FR} 
\end{equation}
If $f=e$, i.e. $G$ is a semidirect product, 
$s$ gives homomorphic embedding $\mathbb{Z}_2 \to G$. 
We could  identify  $x_{1}$ with $b:= s(x_1)$ and, as in 
Eq.~\eqref{eq:fundamental}, write $\sigma_{x_{1}}$ as  $\sigma_{b}$. 
In what follows, we exhibit some examples of the group extension $G$ 
of $\mathbb{Z}_2$ by an abelian kernel $H$.

\subsubsection{Example.1 \quad $H=\mathbb{Z}_{N}= \langle a \rangle $}

Here we take $N = p^e$ with a prime number $p$. 

\begin{itemize}

\item $\sigma_{x_{1}}(h) = h$

Any $f = a^{k}\; (k\in \mathbb{Z})$ satisfy 
the cocycle condition~\eqref{eq:cocycle2}. 
The equivalence relation \eqref{eq:equiv} becomes 
$f \sim {\alpha}^{2}f \sim {\alpha}^{4}f \sim \cdots$. 
Thus for odd $p$ all of $f$'s are equivalent to $e$, 
which means that the group $G$ is the direct product group 
$\mathbb{Z}_{N} \times \mathbb{Z}_{2}$. 
For $p = 2$ there are two cases, i.e. 
\begin{equation}
\begin{cases}
f = a^{k} \sim e & (k:\text{even}), \\
f = a^{k} \sim a & (k:\text{odd}).
\end{cases}
\end{equation}
The former case leads to $\mathbb{Z}_{N} \times \mathbb{Z}_{2}$, 
while the latter leads to 
\begin{align*}
G &=\bigl \langle a,b \bigm | a^{N} =  b^{2}=e, \ 
                          b a b = a  \bigr \rangle \\
&= \bigl \langle a' \bigm | {a'}^{2N} = e \bigr \rangle 
               = \mathbb{Z}_{2N}.
\end{align*} 

\item $\sigma_{x_{1}}(h) = h^{-1}$

The cocycle condition~\eqref{eq:cocycle2} becomes $f=f^{-1}$, 
whose solutions are $f = e, a^{N/2}$. 
The latter is possible only for $p = 2$. 
The group $G$ is a dihedral group
\begin{align*}
 G &=\bigl \langle a,b \bigm | a^{N} =  b^{2}=e, \; 
           b a b = a^{-1}  \bigr \rangle,
\intertext{or a binary dihedral group}
 G & =\bigl \langle a,b \bigm | a^{N} = e, b^{2}= a^{N/2}, \; 
    b a b = a^{-1}  \bigr \rangle  .
\end{align*}

\item $\sigma_{x_{1}}(h) = h^{\pm r}$ 

Similarly to the above consideration, we find 
\[
 G =\bigl \langle a,b \bigm | a^{N} =  b^{2}=e, \; 
                  b a b = a^{\pm r}  \bigr \rangle.
\]
\end{itemize}

\subsubsection{Example. 2 \quad $H=\mathbb{Z}_{N_1} \times 
   \mathbb{Z}_{N_2} = \langle a_{1} \rangle 
                 \times \langle a_{2} \rangle $}

We show some examples for 
$N_1={p_1}^{e_1}, N_2={p_2}^{e_2} $. 
\begin{itemize}
\item
$A =\bigl( \begin{smallmatrix} 1 & 0 \\ 
                               0 & -1\end{smallmatrix} \bigr)$, 
i.e. \ $\sigma (h_{1},h_{2}) = h_{1}{h_{2}}^{-1}$ \\
If  $f = e$, we have the semidirect product 
\begin{equation*}
 G_{1}  =\bigl \langle a_{1},a_{2},b \bigm | {a_{1}}^{N_{1}} = 
                {a_{2}}^{N_{2}}  =  b^{2}= e,\; 
                b\, a_{1} \,b = a_{1},  \; 
                b\, a_{2} \,b ={ a_{2}}^{-1}\bigr \rangle. 
\end{equation*}
When $p_i = 2$ $(i = 1, \ 2)$, $f ={a_i}^{N_i}/2$ is allowed. 
Furthermore, when $p_1 = p_2 = 2$, 
$f = {a_{1}}^{N_{1}/2}{ a_{2}}^{N_{2}/2}$ is also allowed. 
Correspondingly, we have 
\begin{align*}
 G_{2} & =\bigl \langle a_{1},a_{2},b \bigm 
   | {a_{1}}^{N_{1}} = {a_{2}}^{N_{2}}  = e,\; 
       b^{2}= {a_{2}}^{N_{2}/2}, \;b\, a_{1} \,b = a_{1}, 
             b\, a_{2} \,b ={ a_{2}}^{-1}\bigr \rangle,  \\
 G_{3} & =\bigl \langle a_{1},a_{2},b \bigm 
   | {a_{1}}^{N_{1}} = {a_{2}}^{N_{2}}  = e,\; 
       b^{2}= {a_{1}}^{N_{1}/2},\; b\, a_{1} \,b = a_{1},   
             b\, a_{2} \,b ={ a_{2}}^{-1}\bigr \rangle,   \\
 G_{4} & =\bigl \langle a_{1},a_{2},b \bigm 
  | {a_{1}}^{N_{1}} = {a_{2}}^{N_{2}}  = e,\; 
       b^{2}= {a_{1}}^{N_{1}/2}{ a_{2}}^{N_{2}/2},\; 
             b\, a_{1} \,b = a_{1},   b\, a_{2} 
                   \,b ={ a_{2}}^{-1}\bigr \rangle. 
\end{align*}

\item
$ A =\bigl( \begin{smallmatrix} 0 & 1 \\ 
                                1 & 0\end{smallmatrix} \bigr)$ ,
i.e. \ $\sigma ({a_{1}}^{k_{1}},{a_{2}}^{k_{2}} ) = 
            {a_{1}}^{k_{2}}{a_{2}}^{k_{1}}$ 
\begin{equation*}
 G  =\bigl \langle a_{1},a_{2},b \bigm 
   | {a_{1}}^{N_{1}} = {a_{2}}^{N_{2}}  =  b^{2}= e, \; 
      b\, a_{1} \,b = a_{2},  \; 
      b\, a_{2} \,b= a_{1}\bigr \rangle.  
\end{equation*}

\item
$ A =\bigl( \begin{smallmatrix} 1 & 0 \\ 
                                1 & -1\end{smallmatrix} \bigr)$,
i.e. \ $\sigma ({a_{1}}^{k_{1}},{a_{2}}^{k_{2}} ) = 
            {a_{1}}^{k_{1}}{a_{2}}^{k_{1} - k_{2}}$ 
\begin{equation*}
 G  =\bigl \langle a_{1},a_{2},b \bigm 
   | {a_{1}}^{N_{1}} = {a_{2}}^{N_{2}}  =  b^{2}= e,\; 
      b\, a_{1} \,b = a_{1\,}a_{2},  \; 
      b\, a_{2} \,b = { a_{2}}^{-1}\bigr \rangle.  
\end{equation*}

\end{itemize}

\vspace{5mm}



\end{document}